\documentclass[a4paper,fleqn]{cas-sc}

\usepackage[numbers]{natbib}
\usepackage{amsthm}
\let\printorcid\relax 
\usepackage{float}

\def\tsc#1{\csdef{#1}{\textsc{\lowercase{#1}}\xspace}}
\tsc{WGM}
\tsc{QE}

\newcommand{\mathsym}[1]{}
\newcommand{\unicode}[1]{}
\newcommand{\bn}{\mathbf n}
\newcommand{\bx}{\mathbf x}
\newcommand{\wfh}{W_{\text{\begin{tiny}FH\end{tiny}}}}
\newcommand{\wof}{W_{{\begin{small}\text{OF}\end{small}}}}
\newcommand{\cd}{c_{\text{\begin{tiny}DSCG\end{tiny}}}}
\newcommand{\cp}{c_{\text{\begin{tiny}PEG\end{tiny}}}}
\newcommand{\ca}{\chi_{\text{\begin{tiny}APP\end{tiny}}}}

\newcommand{\bm}{\mathbf m}

\newcommand{\bp}{\mathbf p}

\newcommand{\bu}{\mathbf u}

\newcommand{\wh}{W_{\begin{tiny}	{\text{Hex}}	\end{tiny}}}
\newcommand{\Vol}{\begin{small}\text{Vol}\end{small}}

\graphicspath{ {./AggPhen_Figures/} }
\newtheorem{theorem}{Theorem}
\newtheorem{lemma}{Lemma}
\newtheorem{corollary}{Corollary}
\newtheorem*{remark}{Remark}

\begin{document}
\let\WriteBookmarks\relax
\def\floatpagepagefraction{1}
\def\textpagefraction{.001}

\shorttitle{Aggregation in Chromonic Liquid Crystals}    
\shortauthors{Lidia Mrad \it{et al.}}  

\title [mode = title]{Aggregation Phenomena in Lyotropic Chromonic Liquid Crystals}  

\author[1]{Lidia Mrad}
\affiliation[1]{organization={Department of Mathematics and Statistics},
            addressline={Mount Holyoke College}, 
            city={South Hadley},
            postcode={01075}, 
            state={MA},
            country={USA},
            email=\:{lmrad@mtholyoke.edu}}  
\cormark[1]
\cortext[cor1]{Corresponding author}

\author[2]{Longhua Zhao}
\affiliation[2]{organization={Department of Mathematics, Applied Mathematics, and Statistics},
            addressline={Case Western Reserve University}, 
            city={Cleveland},
            postcode={44106}, 
            state={OH},
            country={USA},
         email={\:longhua.zhao@case.edu}}    

\author[3]{Malena I. Espa\~nol}
\affiliation[3]{organization={School of Mathematical and Statistical Sciences},
            addressline={Arizona State University}, 
            city={Tempe},
            postcode={85281}, 
            state={AZ},
            country={USA},
          email={\:malena.espanol@asu.edu}}

\author[4]{Ling Xu}
\affiliation[4]{organization={Department of Mathematics and Statistics},
            addressline={North Carolina A\&T State University}, 
            city={Greensboro},
            postcode={27411}, 
            state={NC},
            country={USA},
            email={\:lxu@ncat.edu}}

\author[5]{M.~Carme Calderer}
\affiliation[5]{organization={School of Mathematics},
            addressline={University of Minnesota}, 
            city={Minneapolis},
            postcode={55455}, 
            state={MN},
            country={USA},
            email={\:mcc@math.umn.edu}}

\begin{abstract}
We study the aggregation phenomenon in lyotropic chromonic liquid crystals as the molecular concentration changes and  condensing agents  are added into the system. Using properties of the critical points of the Oseen-Frank energy of a nematic liquid crystal, combined with the geometric constraints  of the hexagonal columnar chromonic phases, we show that the minimizers of the total energy are topologically equivalent to tori, in agreement with available  experimental evidence on chromonic liquid crystals and DNA condensates, in viral capsids as well as in free solution. We model the system as bi-phasic, consisting of liquid crystal molecules and water, and postulate the total energy as the sum of the Flory-Huggins energy of mixing together with  the bending and surface tension contributions of the liquid crystal. Two types of problems are considered, one related to finding the optimal shape of a torus, once the phase separation has occurred, and the second one that models the conditions leading to molecular aggregation. This work follows recent experimental investigations, but without addressing the topological properties of the toroidal nuclei observed and focusing on how the liquid crystal order competes with the aggregation phenomenon. 
\end{abstract}


\maketitle

\section{Introduction} 
\label{sec:Intro}

Lyotropic  chromonic liquid crystals (LCLCs) consist of disk-like molecules that form columnar aggregates in water, and whose axes tend to align along a preferred direction, constituting the nematic phase. 
The length of the columnar aggregates and their orientational order increase with the molecular concentration. Upon reaching a high concentration threshold,   the increased packing density of the columns leads to a  cross sectional lattice 
structure, with the new phase being dubbed columnar or hexagonal. 
Chromonic liquid crystals distinguish themselves from the thermotropic calamitic class made of rod-like molecules, that achieve increasingly ordered  configurations upon lowering the temperature. These are the liquid crystals  usually  found in display devices.
In the chromonic models, the axis of the cylinders corresponds to the molecular direction of rod-like molecules of calamitic liquid crystals, whose average gives rise to the  unit director field $\bn $ of the macroscopic theory. 
In this paper, we aim at modeling the experimentally observed reorganization and aggregation phenomena that occur  upon adding osmolite molecules in the  columnar phase of the material. The initial disruption of the columnar structure observed upon adding the osmolite, disodium cromoglycate (DSCG) in our case, gives rise to a reorganization of the material into toroidal clusters, presenting topological features such as corners and facets \cite{koizumi2022}.  Alternatively, in the presence of concentrating agents such as polyethylene glycol (PEG ) or spermine, the system also changes structure forming toroidal clusters embedded in their own isotropic media. It is worth noting that such phenomenon is not unique to material at the microscale, but rather extends to the nanoscale. DNA in condensed states also form columnar hexagonal chromonic liquid crystal phases \cite{livolantThese, livolant1980double, livolant1986liquid} and give rise to toroidal-type aggregates \cite{Hud2001, hud2005}. Aggregate states can be found in free solution with condensing agents as well as in bacteriophage viruses, those that infect bacteria (\cite{hiltner2021chromonic} and references therein).

The present  work involves two main modeling components: prediction of the  size and geometry of smooth toroids and a study of the aggregation process, from the molecular components in water to the aggregates. The latter makes use of the Flory-Huggins theory combined with the Oseen-Frank model of nematics, taking into account  the range of parameters of the chromonic phase. 

\begin{figure}
\begin{center}
\includegraphics[width = 0.5 \textwidth]{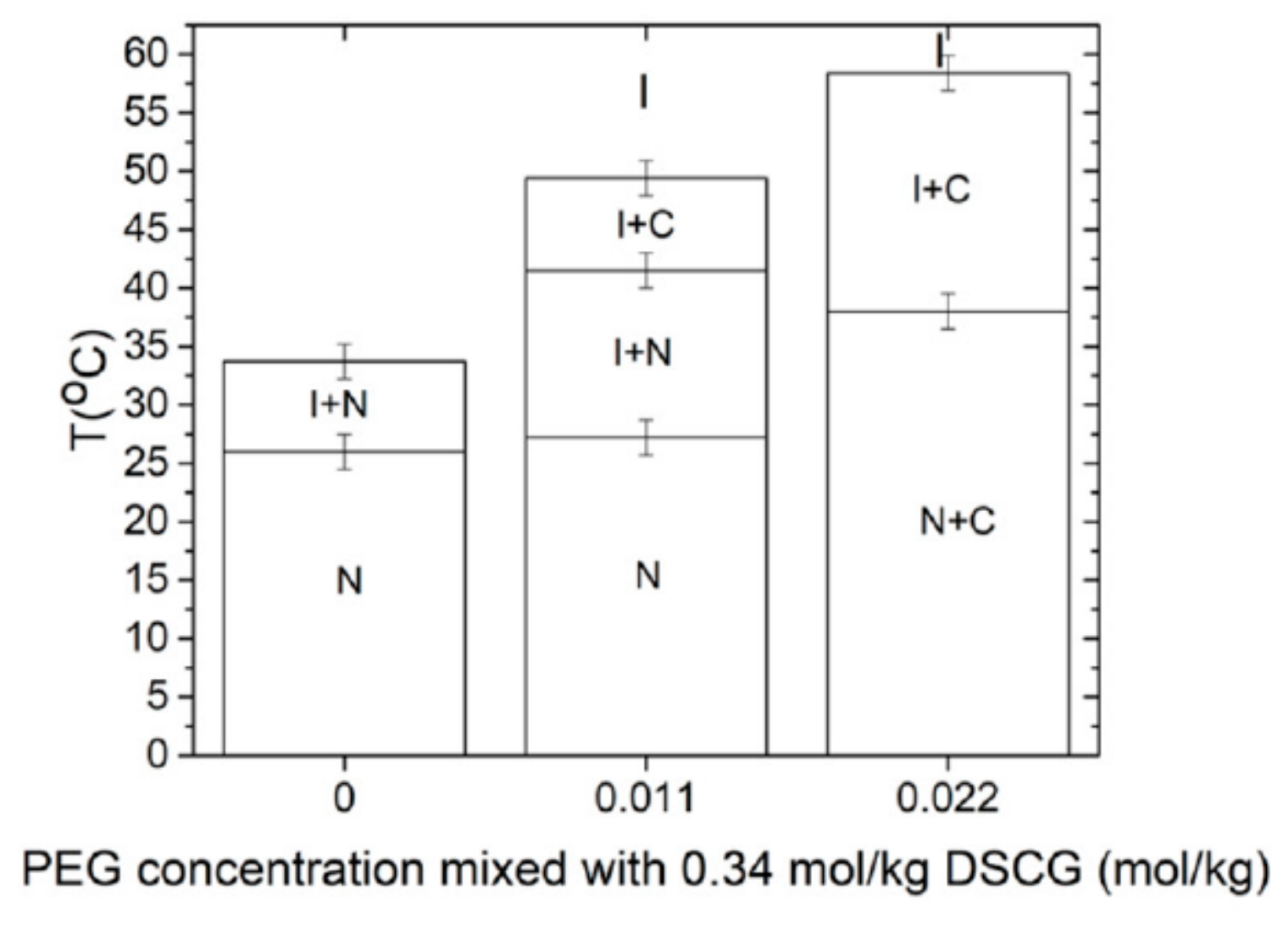}\hspace{0.2in} \includegraphics[width = 0.44 \textwidth]{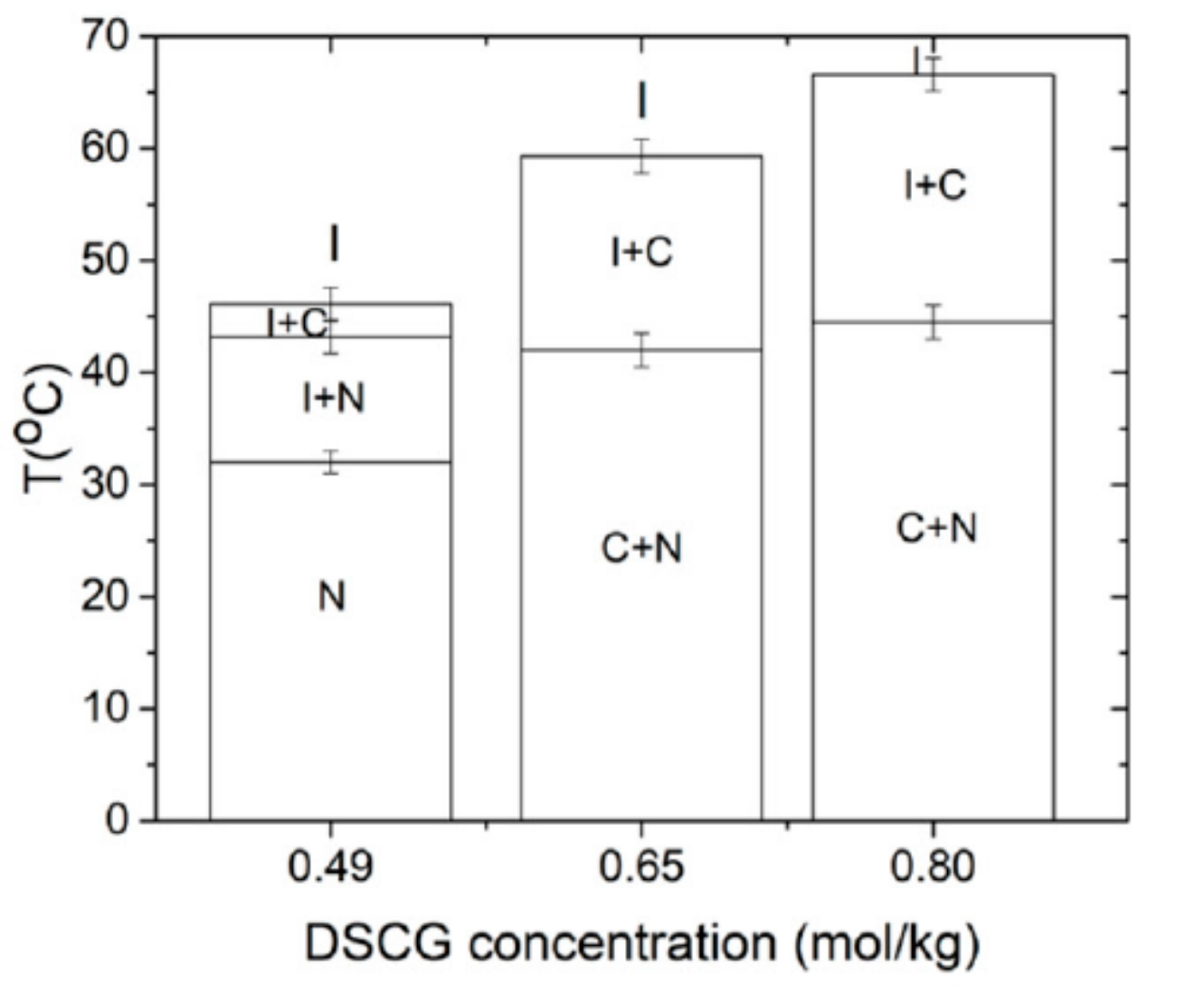}
\caption{Phase coexistence for several PEG concentrations mixed with 0.34 mol/Kg DSCG  (left). Phase coexistence for several DSCG concentrations in mol/Kg (right). We look for toroidal clusters in the regions of I+C coexistence.  (Figures are reprinted from \cite{koizumi2019}.)}
\end{center} 
\label{phase-coexistence}
\end{figure}

Relevant to our work is how phases coexist in chromonic liquid crystals. A detailed phase diagram is presented in \cite{koizumi2019} (see Figure \ref{phase-coexistence}). It shows that when cooling down the material from the isotropic phase (I), the nucleation of the  nematic phase (N) is observed. This regime is characterized by the formation of tactoid domains, where regions of the isotropic phase appear in the nematic bulk \cite{golovaty2020phase}.   The coexistence I+N region transforms into the homogeneous N phase around room temperature. Polarizing microscopy textures suggest that addition of PEG to 0.34 mol/kg of DSCG causes the appearance of the C phase in coexistence with the I phase. (In 0.34 mol/kg DSCG without any PEG, the C phase does not form.)
At around $46^\circ$C, the I+C coexistence region is narrow, less than $5^\circ$C for a 0.49 mol/kg DSCG solution, but expands
as the concentration of DSCG increases (see Figure \ref{phase-coexistence}, right). This work focuses on the I+C coexistence region, for concentration values of DSCG and PEG close to the ones shown in Figure \ref{phase-coexistence}.

Since defects do not play a direct role in the phenomena that we  study, we model the nematic liquid crystal according to the Oseen-Frank energy, quadratic on gradients of the director field $\bn$. This energy together with the elastic energy of the crystalline cross section form the total energy of the columnar hexagonal liquid crystal, subject to relevant geometric constraints; in the case that we model a free boundary aggregate, the surface energy of such a domain is also taken into account. The latter,  together with the elastic cross-sectional energy, provide the material with the necessary cohesion to sustain the columnar hexagonal phase. However, in modeling nematic clusters, one often neglects the elastic energy of the cross section \cite{park2011condensation, lavrentovich2010}. 
The observation of the experimentally obtained toroidal shapes motivates the following modeling assumptions. The aggregates present edges and corners, that, in related work, are accounted for by the anisotropy of the surface tension  \cite{angenent1989multiphase}. Here, we consider the surface energy to be isotropic leading to smooth domains.  Also,  the observed domains are not simply connected,  otherwise a defect core, with a large energy penalty, would be present. This is consistent with minimizing the energy within a class of tori.  Furthermore,  the fact that the liquid crystal molecules are found to be   tangent to the  surface of the aggregate,  motivates us to assume that the latter is a domain   enveloping  families of curves with tangent field $\bn$.
These considerations, together with the conclusions of Marris theorem on the critical points of the Oseen-Frank energy \cite{Marris1978}, determine two types of critical points as relevant to the optimal shapes: $\bn$ arranged either in straight lines or in concentric circles (Theorem \ref{Marris}). 

In our analysis, we scale the experimental domain to a subdomain $\mathcal B$ containing a single torus. We assume that $\mathcal B$ contains a binary mixture of water and DSCG molecules. We assign to $\mathcal B$ a total energy consisting of the Flory-Huggins energy of mixing together with the previously described 
chromonic energy of the toroidal domain \cite{brochard1984phase, flory1979theory-2, flory1953principles, flory1984molecular, flory1979theory, soule2012modelling}. The first energy involves the entropic contributions of each separate component and their energy of interaction, whose strength is represented by the Flory parameter $\chi$.  The torus aggregate forms in the phase separation regime  of such an energy.  However, a scaling analysis reveals several orders of magnitude of dominance of the Flory-Huggins energy with respect to the Oseen-Frank and surface energy terms, in the case that the liquid crystal bending constant $\kappa_3$ is taken to be that of the liquid crystal DSCG \cite{koizumi2019}.  In order to solve such a dilemma, we observe that, in order for the cylinders to  close into tori, it is essential to take into account the energy of interaction of the cylindrical basis. The latter is not accounted for in the total energy as previously described. Rather than explicitly incorporating it into the combined Flory-Huggins and liquid crystal energy form, we treat its contribution as that of an intrinsic bending, allowing us to replace $\kappa_3$ with its effective value (subsection \ref{sec:effective}).   

Motivated by the experiments exhibiting toroidal droplets \cite{koizumi2022}, we perform two types of analysis. The first one corresponds to the case with no PEG added to the system, where increasing the DSCG concentration resulted in observing larger toroidal aggregates. Such a system can be effectively studied through a total energy comprising only of the Oseen-Frank and surface energy terms. Our semi-analytical and numerical results of the minimization of this energy show agreement with the observations of enlarged aggregates corresponding to higher chromonic concentration. The second type of analysis considers the addition of PEG to the system, where increasing the PEG concentration while fixing the DSCG concentration appears to favor similar clustering to the previous case. For this case, the mixing energy of Flory-Huggins is added to the total energy, and our numerical minimization of this total energy shows consistency with toroidal observations.

This article is organized as follows. In Section~\ref{sec:Energy}, we present a survey on chromonic liquid crystals, the energy forms used in their analysis, justifying our focus on  the toroidal shapes. In Section~\ref{sec:mathmodel},  we introduce our mathematical model including the assumptions on the geometry, the scaling analysis, and the parameters needed for the model. Section~\ref{sec:optimal} is devoted to finding the optimal torus shape that minimizes the combination of bending and surface energy. An outcome of the analysis is finding the dependence of the energy ratios on the concentration of DSCG. This approach is consistent with assuming that phase separation has already taken place and so we neglect the forces causing it. In Section~\ref{sec:FH}, the phase separation is analyzed by including the Flory-Huggins energy of binary systems. This reveals the relevant role of the anisotropic interaction between chromonic molecules and the dependence of the Flory parameter $\chi$ on PEG concentration emerges.  In Section~\ref{sec:conclusion}, we present our conclusions. 

We conclude the introduction by recalling the Oseen-Frank energy of nematic liquid crystals.  Let $\bn\in\mathcal S^2$ denote the unit vector field. 
The total energy of a nematic liquid crystal occupying a domain $\mathcal U\subset {\mathbb R}^3$ is  given by \begin{align}
E(\bn)=\int_{\mathcal U} \wof(\bn, \nabla\bn)\,d\bx
\end{align}
with
\begin{align}
2\wof(\bn, \nabla\bn)=\kappa_1(\nabla\cdot\bn)^2+ \kappa_2(\bn\cdot\nabla\times\bn+\tau)^2+ \kappa_3|
\bn\times\nabla\times\bn|^2+(\kappa_2+\kappa_4)\left(\mbox{tr}(\nabla\bn)^2-(\nabla\cdot\bn)^2\right),
\label{OF}
\end{align}
where the Frank elasticity constants $\kappa_1, \kappa_2, \kappa_3$, and $\kappa_4$ satisfy the  inequalities
	\begin{equation}
\kappa_1>0, \, \kappa_2>0,\, \kappa_3>0, \, \kappa_2\geq |\kappa_4|, \, 2\kappa_1\geq \kappa_2+\kappa_4. \label{elasticity-constants}
\end{equation} 

When working with the Oseen-Frank energy, as a  simplification, we assume that the saddle-splay constant, $\kappa_2+\kappa_4$, is zero. The importance of this constant in similar settings, however, has been recently underscored \cite{koning2014}. For example, a large saddle-splay, along with the right boundary conditions on cylinders with chromonic liquid crystals, would lead to a double-twist configuration \cite{davidson2015}. We keep our assumption since no twists, or other defects, are observed for the experiments we study.
\section{Modeling the hexagonal or columnar phase}\label{sec:Energy} 

Formation of cylindrical aggregates, with  the axial direction corresponding to that of molecular rigid rods in calamitic liquid crystals, is prompted by the fact that DSCG are hydrophobic on their broad side and hydrophilic on the lateral surface. At a threshold concentration, these cylinders from the nematic phase aggregate in groups of six, leading to the columnar or hexagonal phase, with two-dimensional crystal order on planes perpendicular to the axial direction. These then tend to cluster producing toroidal shapes. Additionally, the presence of concentrating agents, such as PEG and salt, within specific concentration regimes produces toroidal-type clustering. The competition between surface and bulk energy, mostly
associated with bending, determines the shape of the clusters. These show an overall lengthening of the
torus as a remedy to defray the energy cost of bending as the inner torus region closes down to reduce surface energy. Typical cluster dimensions are in the order of 30 to 100 $\mu m$ \cite{koizumi2022}.

We now present several approaches for the modeling of the columnar chromonic phase. First, we note that 
the geometry of the hexagonal phase is represented by a triple of orthonormal vectors $\{\bn, \bm, \bp\}$, with $\bn$ describing the average alignment of the axes of the columnar molecular  aggregates, the {\it liquid} direction, and $\bm$
 and $\bp$ the lattice vectors of the plane {\it solid} lattice.  The energy density consists of the sum of the Oseen-Frank energy of the  nematic liquid crystal (appropriate to problems where defects do not play a main role) and that of a two-dimensional crystal, together with the required geometric constraints. 
The two-dimensional crystal energy penalizes deformation of the cross section resulting in a cohesive effect on the material.

Letting $\Omega\subset {\mathbb R}^3$ denote the region occupied by the hexagonal chromonic liquid crystal, the total energy is 
\begin{align}
E_{chr} & = \int_\Omega \left\{\wof(\bn, \nabla\bn) + \wh(\nabla\bu) \right\} d\bx,\label{basic-chrom-energy-v0} \\
&\text{subject to }\nabla\cdot\bn = 0\quad \text{in } \Omega, \text{ where } \bu := u(\bx) \bm+ v(\bx)\bp,\nonumber
\end{align}
with $\bu$ denoting displacements of the cross section.  Dirichlet (or weak anchoring) boundary conditions on $\bn$ and $\bu$ also need to be prescribed. 
 Well-known forms of $\wh$  
can be found in the works by de Gennes and Kl\'eman \cite{degennes1993, kleman1980}, appropriate to small displacements and bending deformations,  followed  by the nonlinear elastic expressions by Oswald and Pieransky \cite{oswald2005smectic}. The role of $\wh$ becomes more prominent in applications to DNA clustering. For example, a form of  $\wh$ is proposed in \cite{hiltner2021chromonic},  that takes into account the elastic material being made of filaments. 
In this case, it is assumed that 
\begin{flalign}
\wh(\bn, \bm, \bp)= B|\nabla (\bm-\bp)|^2 +
 C |\nabla (\bm+\bp)|^2, \label{wh}\\
\text{with }\bm = \bn \times \bp, \: \bn\cdot\bp=0, \: |\bm|=|\bn|=|\bp|=1,\label{ortho}
\end{flalign}
where the constants $B>0$ and $C>0$ represent shear and compressible moduli, respectively. An alternate free energy expression to account for the cohesiveness of the columnar phase has been used in studies of
DNA packing in \cite{klug2003director}.

Alternatively, in studies of free boundary structures, such as the tori clusters addressed in this work, the hexagonal energy is replaced by a surface tension aimed at penalizing any increases in area and hence providing an adhesive role. Suppose now that the domain $\Omega$  has a  free boundary $\partial \Omega$. In this case,  we set up the total energy as 
\begin{flalign}
E_{chr} &= \int_\Omega\wof(\bn, \nabla\bn) d\bx +  \sigma\mbox{Area}(\partial \Omega),
\label{basic-chrom-energy} \\
&\text{ subject to }|\bn| = 1,\quad\nabla\cdot\bn = 0 \text{ in }\Omega, \quad \bn\cdot\nabla\times\bn = 0 \text{ in }\Omega,\nonumber \\
&\quad\bn\cdot\boldsymbol \nu =0 \quad \text {on}\quad \partial\Omega,\quad\Vol(\Omega) = V_0, \nonumber
\end{flalign}
with $V_0$ constant. The positive constants $\kappa_3$ and $\sigma$ denote the bending and surface tension moduli, respectively. Here $\boldsymbol \nu$ denotes the unit vector field perpendicular to $\partial\Omega.$

The role of the constraints, $\nabla\cdot\bn = 0$ and $\bn\cdot\nabla\times\bn = 0$, is to express the large resistance to splay and twist deformation of the molecules, respectively. Assuming dislocations do not occur, the same number of columns entering a cross section should also exit it. This leads to the assumption that splay is zero, that is , $\nabla\cdot\bn = 0$, since in the case of the hexagonal columnar phase, nonzero splay would allow for deviations from the lattice structure. On the other hand, it is assumed that there is no internal twist because of its incompatibility with the two-dimensional lattice order in planes perpendicular to the director. This results in setting $\bn\cdot\nabla\times\bn = 0$. An approximation to this constraint model can also be achieved through relaxation. That is, alternatively, one can take into account the dominance of the splay and twist constants over the bending one by requiring
        \begin{align}
        \kappa_1, \kappa_2 \gg \kappa_3.\label{constraint-relaxation}
        \end{align}
However, in our work, we assume the above constraints instead of the relaxation.

In order to guarantee the existence of minimizers, we recall a fundamental result in the analysis of energy minimization of nematic liquid crystals. 
\begin{theorem}\label{HKL}~\cite{hardt1986} Let $\mathcal U \subset {\mathbb R}^3$ be an open and bounded set, with Lipschitz boundary $\partial\mathcal U$. 
	Suppose that the Frank constants satisfy the  inequalities (\ref{elasticity-constants}).
Let  the admissible set 
$$\mathcal A(\bn_0)=\{\bn\in H^1(\mathcal U, {\mathcal S}^2): \text {trace of }\, \bn=\bn_0 \}$$ be nonempty. 
	Then, for any Lipschitz function $\bn_0\colon \partial \mathcal U \rightarrow {\mathcal S}^2 $, the functional $E(\bn)=\int_{\mathcal U} \wof(\bn, \nabla\bn)\,d\bx$ admits a minimizer in $\mathcal A(\bn_0)$.
	Furthermore, if $\bn$ is a minimizer of $E(\cdot)$, then $\bn $ is analytic on $\mathcal U/Z$ for some relatively closed subset $Z$ of $\mathcal U$ which has one dimensional Hausdorff measure zero.
\end{theorem}
Note that the inequalities (\ref{elasticity-constants}) are needed to guarantee the coercivity of $E(\bn)$. In the case of $E(\bn)=\int_{\mathcal U}\{\wof(\bn, \nabla\bn)+\wh(\nabla\bm, \nabla\bp)\}\,d\bx,$ steps analogous to those that lead to the conclusion of Theorem~\ref{HKL}, in the case that $\bp$ is a prescribed constant vector, also prove existence and partial regularity of minimizers of the energy 
$E(\bn)$ in the admissible set
$\mathcal A(\bn_0,\bm_0)= \{\bn, \bm\in H^1(\mathcal U, {\mathcal S}^2): \text {trace of}\, \bn=\bn_0, \,  \text {trace of}\,\bm=\,\bm_0, \, \text {subject to}\, (\ref{ortho}) \}$, for a given pair of unit vector fields, $\bm_0, \bn_0 \in H^1(\mathcal U)$, satisfying $\bm_0\cdot\bn_0=0$.

\subsection {Critical points of the Oseen-Frank energy}
These fields, known as universal solutions, 
 were first reported by Ericksen \cite{ericksen1967}, for a general class of nematic free energy densities
$W(\bn, \nabla\bn)$ satisfying the properties of frame-indifference and invariance with respect to the transformation
$\bn \to -\bn$. These solutions consist of families of vector lines, including  either parallel straight lines, or lines arranged in
a uniform twist along an axis with constant directions on the perpendicular planes, lines orthogonal to a
family of concentric spheres, or coaxial cylinders.
Furthermore, for the Oseen-Frank energy, Marris classifies all the critical points in the following theorem: 

\begin{theorem}\cite{Marris1978}\label{Marris} \,
The only possible universal equilibrium configurations for liquid
crystals in the nematic state, with $W$ quadratic in $\text{grad}\, \bn$, are the following:
\begin{enumerate}
    \item The (straight)  vector-lines of $\bn$ that (also)  comprise the rectilinear uniplanar field $\bn=(\sin\tau z, \cos\tau z, 0)$, with $\tau$ constant.
\item The vector-lines of $\bn$  are the orthogonal trajectories of a family of parallel planes, a family of concentric spheres, and a family of concentric circular cylinders.
\item The vector-lines of $\bn$  are either concentric circles or a family of circles
capable of being intersected orthogonally by the members of a second family of
circles.
\end{enumerate}
\end{theorem}

\subsection {Free boundary minimizers of the chromonic energy (\ref{basic-chrom-energy})  }
Although the previous theorem gives the full list of critical points in the bulk, we are looking for energy minimizers in three-dimensional,  not simply connected  domains of prescribed volume, bounded by  (smooth) developable surfaces, that is, with zero Gaussian curvature. Moreover, we will ignore the edges and corners shown in the experimental shapes. The optimal shapes, both in the case of chromonic liquid crystals and DNA clusters, are shown to be toroidal. It is well known that the torus has a metric under which it is developable, which can be embedded in the three-dimensional space by the Nash embedding theorem \cite{borrelli2012flat}. Furthermore, we restrict ourselves to the case when the material is nonchiral (that is, when $\tau=0$), and, accordingly,  look for minimizers in axisymmetric domains. With these observations, we proceed to the construction of free boundary minimizers of the energy \eqref{basic-chrom-energy}.  

 Let us consider three-dimensional domains, with prescribed volume $V_0$, bounded by smooth or piecewise smooth surfaces that are envelops of either family of vector fields, within the class of critical points given in Theorem \ref{Marris}:
\smallskip{}{}

\noindent {\bf Class 1.\, } Straight lines (that without loss of generality, can be taken along the $z$-axis)  contained in a cylinder of volume $V_0$.
    \smallskip{}
    
    \noindent 
   {\bf Class 2.\,}   Concentric circles filling a torus of volume $V_0$.

Determining the optimal shapes and the corresponding estimates on the coefficients of the energy is done in Section \ref{sec:optimal}. Motivated by the experimental observations reported in \cite{koizumi2022} which are relevant to our current study, our focus is on the second class of critical points obtained above. In subsection  \ref{CylindertoTorus}, we calculate the energy needed for a cylinder to rearrange itself into a torus.

\medskip

\section{Mathematical model}\label{sec:mathmodel}
There are two main approaches to the  problem of finding the optimal torus. In the simpler approach, given an unknown torus domain, one looks for the optimal shape that minimizes the combined bending and surface energies. The second approach addresses how a toroidal domain nucleates from the isotropic phase of the material. In such a case, in addition to the energy needed to produce the phase separation, with the resulting clustered domain being in the hexagonal columnar phase (C),  it is also necessary to account for the  bending energy required to rearrange the column structures into a  torus, while subject to a surface energy penalty. 

Following the second approach, we  set up the total energy of the system consisting of the bending and the surface energies, together with the Flory-Huggins energy of mixing. The latter, originally developed for mixtures of polymers and solvents, is used in modeling binary systems where the entropy of the components competes with their chemical interactions. 
Let $\mathcal B \subset{\mathbb R}^3,$ a sphere of radius $R_0$,  denote a domain encompassing a single torus $\Omega$. We will take $R_0$ to be a quantity of the same order of magnitude as the experimentally observed torus. The variables of the problem consist of the unit director field $\bn$ of the nematic liquid crystal and the volume fraction $\phi$ of chromonic molecules, in addition to the geometrical dimensions of the torus. We assume that the system is saturated, that is, it contains only liquid crystal molecules and water, the volume fraction of the latter being then $1-\phi$.

In order to simplify the presentation of our mathematical model, we will make explicit the dependence of the volume fraction $\phi$ on $\bx$, and write $\phi(\bx),$  $\bx\in \mathcal B$, to denote the volume fraction of chromonic molecules at that point. Consequently, $1-\phi(\bx)$ denotes the volume fraction of water at that same point. 

The total energy of the system consists of the Flory-Huggins energy of mixing \cite{de1979scaling} together with the bending and surface energies of a torus, which we represent by $\Omega\subset {\mathbb R}^3$:
\begin{equation}E(\mathbf{n}, \phi, \Omega) = E_b(\mathbf{n},\phi,\Omega)+E_s(\Omega)+E_{FH}(\phi,\Omega),\label{total_energy}\end{equation}

where
\begin{eqnarray}
	E_b&=&\int _{\Omega }\phi \left(\kappa _3|(\nabla \times \mathbf{n})\times \mathbf{n}|^2\right)d\mathbf{x}\label{bending_energy}, \\
	E_s&=&\int _{\partial \Omega }\sigma _0 ds\label{surface_energy}, \mbox{ and}\\
	E_{FH}&=& \frac{R T}{V_m}\int_{\mathcal B}\left[\frac{1 }{N_1}\phi \ln\left(\phi\right)+\frac{1}{N_2}\left(1-\phi \right) \ln\left(1-\phi\right)+\chi  \phi\left(1-\phi\right)\right]d\mathbf{x}\nonumber\\
	 &:= &\int_{\mathcal B} \wfh(\phi(\bx))\,d\bx, 
	\label{FH_energy}
\end{eqnarray}
with $\kappa _3$ being the bending Frank elastic constant of the torus, $\sigma_0$ the surface tension coefficient, $R$ the ideal gas constant, $T$ the absolute temperature, and $V_{m}$ the molar volume of water. The parameters $N_1$ and $N_2$ denote the number of lattice sites occupied by the solute (liquid crystal) and solvent respectively, and $\chi$ is the dimensionless Flory interaction parameter. In our case, $N_1>>N_2=1$, since basic chromonic units in the isotropic phase are formed by stacking a few hundred individual molecules (see end of subsection \ref{sec:effective} for an explanation of the specific $N_1$ value we take). 

We note that in our definition of the bending energy, (\ref{bending_energy}), we remove the dependency of the Frank constant on the volume fraction, $\phi$, since $\phi$ is one of our model variables. Our choice is based on \cite{zhou2014}, where the authors show that the bending elastic constant for chromonic liquid crystals is proportional to $\phi$. So, in our calculations from now on, we obtain our bending constant $\kappa_3$ through dividing by $\phi$ (see Table \ref{table-parameters}).

The bending energy (\ref{bending_energy}) of an individual torus follows from the Oseen-Frank energy of a nematic liquid crystal subject to the constraints:
\begin{enumerate}
	\item The  director $\mathbf{n}$ is a unit  vector,  $|\mathbf{n}|=1$.
	\item $ \nabla\cdot \mathbf{n}=0$.
	\item $(\nabla \times \mathbf{n})\cdot \mathbf{n}=0$.
\end{enumerate}
The latter two constraints express the fact that the Oseen-Frank energy of the chromonic liquid crystals is bending-dominated, as discussed before. In addition, we have a constraint representing the conservation of the total amount of liquid crystal molecules, that, given a fixed density, is expressed in terms of their volume $c$ as 
 \begin{equation}
\int _\mathcal{B}\phi\,d\mathbf{x}=c.\label{constraint}
\end{equation}
 The goal of our work is to minimize the total energy (\ref{total_energy}-\ref{FH_energy}) subject to the constraint (\ref{constraint}), in the case that $\Omega$ is a torus, and $\bn$ is a prescribed unit director field corresponding to the domain geometry, i.e., $\bn={\mathbf e}_\theta$, the azimutal vector of the cylindrical coordinate system.

\subsection{Scaling, dimensional analysis and effective Frank constants}\label{sec:effective}
 Let $R_0$ denote a characteristic length at the microscopic scale, that is, a quantity of the order of magnitude of the size of an experimentally observed torus, in the order of 10$\mu m$. We first identify the energy scales associated with each of the energy components of (\ref{total_energy}). We have
 \begin{align}
 	E_b&=\kappa _3 R_0\int _{\tilde \Omega }\phi\left(|(\nabla \times \mathbf{n})\times \mathbf{n}|^2\right)d\bar{\mathbf{x}}\label{bending_energy_2},\\
 E_s&=\sigma _0R_0^2\int _{\partial \tilde\Omega } d\bar s\label{surface_energy_2}, \mbox{ and}\\
 E_{FH}&=\frac{4}{3}\pi R_0^3\frac{R T}{ V_{m}}\int _{\bar {\mathcal B}}\left[
 \frac{1}{N_1}\phi \ln\left(\phi\right)+\frac{1}{N_2 }\left(1-\phi \right) \ln\left(1-\phi\right)+\chi  \phi\left(1-\phi\right)\right]d\bar{\mathbf{x}}, \label{FH_energy_2}
  \end{align} 
 where $\bar {\mathcal B}$ denotes the sphere of radius 1. Note that the coefficients of each of the previous integrals have the dimensions of an energy.
 Let us denote the energy scales by
    \begin{equation}
     K=\kappa_3 R_0, \quad S=\sigma_0 R_0^2, \quad F= \frac{4}{3}\pi R_0^3\frac{R T}{ V_{m}}:=\left(\frac{4}{3}\pi R_0^3\right)\nu. \label{energy-scales}
     \end{equation}
     Taking parameter values from Table \ref{table-parameters} gives
     \begin{equation*}
     K= 2.175\cdot 10^{-15}\,\text{J}, \, \nu= 1.261\cdot 10^8 \text{ J/m}^3, \, F=2.282\cdot 10^{-7} \text{ J}.
     \end{equation*}
    To take into account that the liquid phase of the binary mixture is not pure water, but rather the isotropic phase of the mixture, we introduce an effective value for Flory energy scale $\nu_{\begin{small}{\text{eff}}\end{small}} = RT/V_{\begin{small}{\text{eff}}\end{small}},$  where 
    $V_{\begin{small}{\text{eff}}\end{small}}$ is taken to be the mean of $V_m$ and $V_{ml}$, the molar volume of DSCG. This yields $V_{\begin{small}{\text{eff}}\end{small}} = 1.7416 \cdot 10^{-4}$ m$^3$, $\nu_{\begin{small}{\text{eff}}\end{small}}= 1.303 \cdot 10^7\text{ J/m}^3$  and consequently 
    $F_{\begin{small}{\text{eff}}\end{small}}=5.458\cdot 10^{-8} \text{ J}.$
    
   We now focus on the gap in order of magnitude between $K$ and 
    $F_{\begin{small}{\text{eff}}\end{small}}$, that is
    \begin{equation}
        \frac{F_{\begin{small}{\text{eff}}\end{small}}}{K}= 2.509 \cdot 10^7.
    \end{equation}

This is an indication that the bending modulus of the DSCG is not sufficient to provide enough  energy to  bend columnar rods and  form tori. That is, with the bending energy being a small perturbation of the total energy, dominated by Flory-Huggins, only light disruptions of cylindrical shapes could be expected.  In order to account for this apparent energy gap, we take into account two additional contributions to bending: 
\begin{enumerate}
    \item The elastic energy density $\wh$ of the columnar cross section of the hexagonal phase (\ref{ortho}). Adding its corresponding energy to the total energy (\ref{total_energy}), and setting $\bn=\mathbf e_\theta$, it is found that $\wh$  provides a contribution as in (\ref{bending_energy_2}). In particular, this results in an enhanced effective bending modulus  $\kappa_3$, that is, the sum of the DSCG bending constant plus the transverse elasticity modulus of $\wh$.  This is found by the same calculation as in \cite{hiltner2021chromonic} for the columnar phase of viral DNA. In that case, the effective $\kappa_3$ turns out to be 100-1000 times the original bending modulus. In the present case,  lacking information to evaluate the contribution of the transverse elasticity, and also taking into account the likely smaller elasticity modulus of the columnar DSCG phase compared to the (polymer-like) DNA one, we assign to it an order of magnitude of 10. This would result into the estimate
    \begin{equation}
    \frac{F_{\begin{small}{\text{eff}}\end{small}}}{K}= 2.509 \cdot 10^6.
    \end{equation}
    \item The hydrophobic energy of the bases of the cylindrical aggregates forming the hexagonal phase have not, so far, been taken into account. It is this hydrophobic energy that causes a cylinder to close down into a torus. In order to quantify this effect, we interpret it as an intrinsic bending contribution, that, in particular, would result into an effective bending modulus. Although we lack the appropriate information to quantify such a  contribution, we use it as a motivation to allow the effective order of magnitude take values 
     \begin{equation}
    \frac{F_{\begin{small}{\text{eff}}\end{small}}}{K_{\begin{small}{\text{eff}}\end{small}}}= O(1), \dots, O(10^3). \label{effective-eta} 
\end{equation}
In addition to the hydrophobic energy, which is the cost of keeping the open ends of the cylinder open, another type of energy affects the bending modulus, to a lesser degree, and that is the scission energy (the energy needed to cut an aggregate into two). It is the balance of this energy and the entropy gained by producing more open ends that results in the formation of columnar aggregates in the first place. The effect of the scission energy is, in fact, direct on the bending modulus in the following way: $\kappa_3$ is directly proportional to the persistence length $\lambda_p$, the length over which the unit vectors tangential to the aggregates lose their correlations \cite{zhou2012, zhou2014}. Monte Carlo simulations \cite{kuriabova2010} showed that $\lambda_p$ depends on the scission energy $E$ in a linear fashion; $\lambda_p \propto 5.07 + 2.14 E/K_BT$. Therefore, an increase in the scission energy implies an increase in $\kappa_3$. The value of the scission energy for DSCG in the nematic phase has been approximated to be in the range $(8-14) K_BT$ \cite{zhou2014}, though, counterintuitively, lower values of $3.5 K_BT$ have been calculated for columnar phases \cite{joshi2009}.

Other factors that play important roles in the aggregation process itself are \cite{joshi2009}: enthalpy forces based on $\pi-\pi$ interactions between the aromatic cores, inter-aggregate interactions controlled by the excluded volume effect, electrostatic repulsion between ionic groups, etc ... , though these are not unique to the columnar phase which is the one we work with here.

\item An increased (effective) bending modulus would result in 
tori with greater radius $R_1$, and therefore, larger surface areas. This motivates us to also consider an effective surface tension $\sigma_{\begin{small}{\text{eff}}\end{small}}$ so that the quotient satisfies  \begin{equation}
\frac{S_{\begin{small}{\text{eff}}\end{small}}}{K_{\begin{small}
{\text{eff}}\end{small}}}=\frac{S}{K}.\end{equation}
\end{enumerate}

We end this subsection by discussing the value of $N_1$, the number of lattice sites occupied by the DSCG molecules per lattice site. This number is deduced from the aggregate contour length, which is several factors greater than the persistence length $\lambda_p$ mentioned above. This is due to the presence of molecular shift junctions, such as the c- and y-type junctions \cite{zhou2014}. Both lengths vary with temperature, concentration, and presence of the condensing agent. The contour length for DSCG is estimated to be in the range 20-270 nm \cite{zhou2014}, which is equivalent to 60-810 molecules, assuming that the ``thickness'' of the chromonic molecule is around 0.34 nm \cite{zhou2017}. This allows us to consider $N_1$ to be on the order of hundreds, which is the value we use in our numerical simulations in Section \ref{sec:FH}.

{\begin{small}{
\begin{table}[htb!]
\caption{Parameter list of our energy model along with their typical values in the context of chromonic liquid crystals.}
\label{table-parameters}
\centering
 \begin{tabular}{lllllll} \hline 
 	Parameter & Label & Unit & Dim. & Value (SI) & Reference/Formula \\ \hline
 	 Characteristic length & $R_0$    & m 	& 		[L]	&  $10^{-5}$ & \cite{koizumi2022}\\[5pt] 
 	Elastic constant & $\kappa_3$    & J/m 	& 	$[F]$ &	 $25\cdot 10^{-12}/0.115$ &	 \cite{zhou2014}, $\phi=0.115$ \\[5pt]
  Ideal gas constant& $R$    & J/(K.mol) 	& 	$\frac{[F]}{[K][mol]}$ &	 8.314462   & \cite{tiesinga2021}	 \\[5pt] 
 Absolute temperature & $T$    & K	& 	$[K]$ &	 273 &  \cite{tiesinga2021}\\[5pt]
 Boltzmann constant &$K_B$&  J/K &$\frac{[F]}{[K]}$&$1.38\cdot10^{-23}$  & \cite{tiesinga2021} \\[5pt]
Molar volume of water& $V_{m}$    & m$^3$/mol	& 	$\frac{[L]^3}{[mol]}$ &	 $1.8\cdot 10^{-5}$ &	$18\cdot 10^{-3}/10^3$  \\[5pt]
 Molar mass of DSCG& $M_{ml}$    & Kg/mol	& 	$\frac{[M]}{[mol]}$ &	0.512 & \cite{koizumi2019}	 \\[5pt]
 Density of DSCG& $\rho$    & Kg/m$^3$	& 	$\frac{[M]}{[L]^3}$ &	 $1.55\cdot 10^3$ &	 \cite{koizumi2019}	  \\[5pt]
  Molar volume of DSCG& $V_{ml}$    & m$^3$/mol	& 	$\frac{[L]^3}{[mol]}$ &	 $3.3032\cdot 10^{-4}$ & $M_{ml}/\rho$ \\[5pt]
FH-energy  coeff. water phase  & $\nu$   & J/${\text m}^3$  & $\frac{[F]}{[L]^3}$    & $1.261\cdot 10^8$ & $ RT/V_{m}$\\[5pt]
FH-energy  coeff. iso. phase  & $\nu_{\begin{small}{\text{eff}}\end{small}}$    & J/${\text m}^3$  &  $\frac{[F]}{[L]^3}$ & $1.303 \cdot 10^7$ & $RT/V_{\begin{small}{\text{eff}}\end{small}}$\\[5pt] 
 No. solvent mol. units/lattice site        & $N_2$    &	& 	 &	 1 & \\[5pt]
 No. DSCG mol. units/lattice site        & $N_1$  &	& 	 &	 100 & \cite{zhou2014}\\[5pt]
\hline
 \end{tabular}
\end{table}}\end{small}} 

\section{Optimal torus shape}\label{sec:optimal}
\begin{figure}[!ht]
\centering
\includegraphics[width=0.4\textwidth]{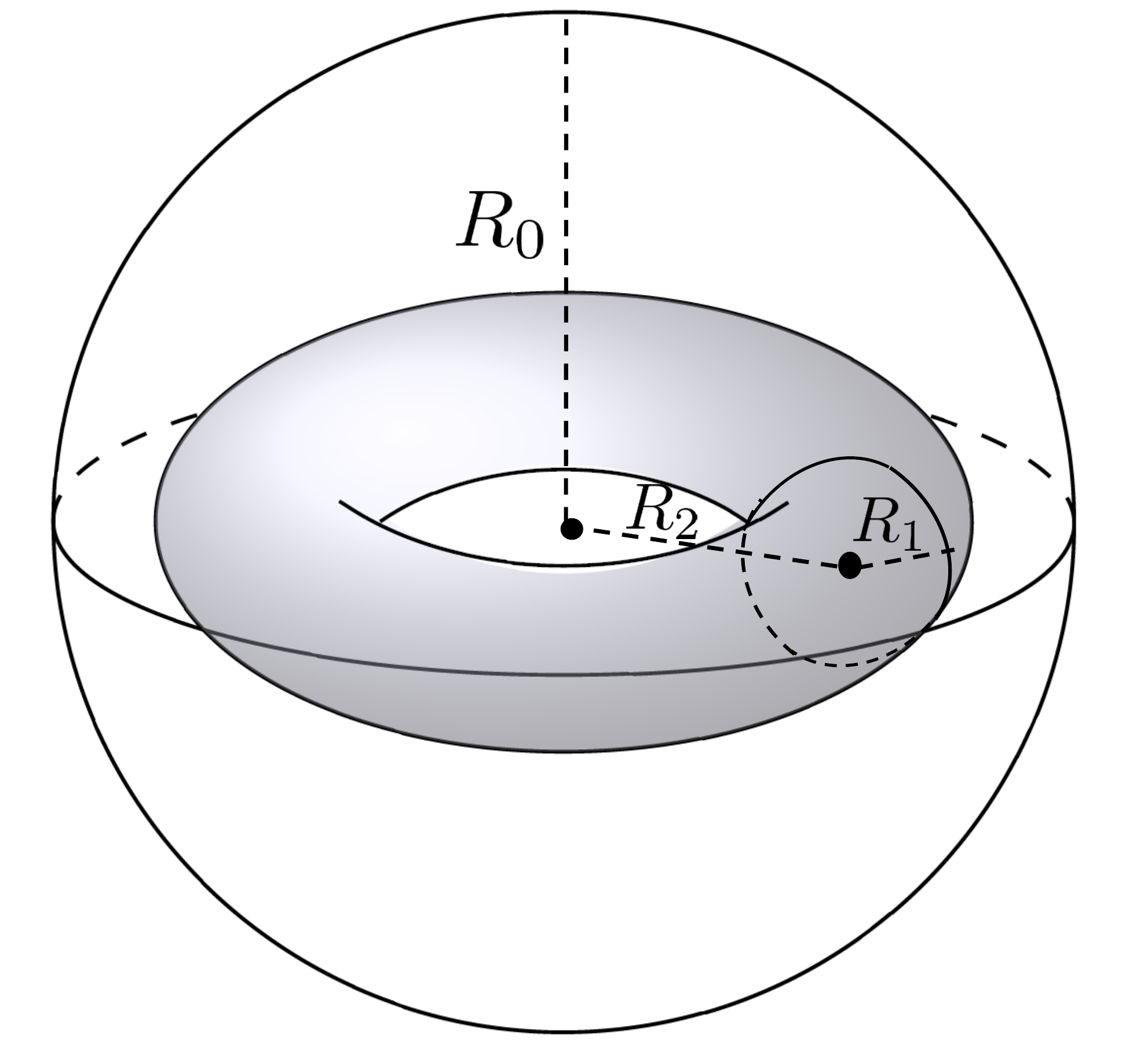}
\caption{
Simplified geometry of the toroidal droplet: a torus with inner radius $R_1$ and outer radius $R_2$ contained in a virtual sphere of typical radius $R_0$.
}\label{figure:simplifiedGeometry.png}
\end{figure}

Guided by experimental observations of tori, we simplify the geometry of the aggregates accordingly. We assume that $\Omega$ denotes a torus 
 with unknown inner and outer radii, $R_1$ and $R_2$, respectively. Furthermore, we assume that the mixture occupies a spherical bath of radius $R_0$. These assumptions on the resulting shape and the bath introduce two geometrical constraints: $0<R_1<R_2$ and $R_1+R_2<R_0$, illustrated in Figure~\ref{figure:simplifiedGeometry.png}.
With this geometrical setup, at any point $(x,y,z)$ inside the torus, the director field is tangential to the columnar rings and thus can be written as $\mathbf{n} = \langle \frac{-y}{\sqrt{x^2+y^2}}, \frac{x}{\sqrt{x^2+y^2}}, 0\rangle$. This guarantees that the first three constraints are satisfied. We assume that DSCG is uniformly mixed inside the torus, with volume fraction $\phi^{in}$, and outside the torus, with volume fraction $\phi^{out}$. With these assumptions, we are able to integrate the bending energy density using spherical coordinates (see Appendix \ref{appendix}). This casts our problem as a minimization problem of the following energy 

\begin{flalign} \label{eq: Torus Energy}
&E\left(R_1,R_2,\phi^{ in},\phi^{ out }\right)=
 4\pi^2 \kappa _3 \phi^{in} \left(R_2-\sqrt{R_2^2-R_1^2}\right)
+ 4 \pi^2\sigma _0  R_1R_2&
 \\&\quad+\nu
\left[\frac{1}{N_1}\phi^{ in} \ln\left(\phi^{ in}\right)+\frac{1}{N_2}\left(1-\phi^{ in}\right) \ln\left(1-\phi^{ in}\right)+\chi 
 \phi^{ in}\left(1-\phi^{ in}\right)\right]\left(2 \pi^2 R_1^2R_2\right)&\nonumber\\
 &\quad
+\nu\left[\frac{1}{N_1}\phi^{ out } \ln\left(\phi^{ out }\right)+\frac{1}{N_2}\left(1-\phi^{ out }\right) \ln\left(1-\phi^{ out }\right)+\chi 
 \phi^{ out }\left(1-\phi^{ out }\right)\right]\left(\frac{4}{3}\pi R_0^3-2 \pi^2 R_1^2R_2\right),&\nonumber
\end{flalign}

subject to the constraints
\begin{align}
	&\quad c=\phi^{ in} \left(2 \pi^2 R_1^2R_2\right)+\phi^{ out }\left[\frac{4}{3}\pi R_0^3-2 \pi^2 R_1^2R_2\right],\\
	&0\leq \phi^{in},\phi^{out} \leq 1,\: 0<R_1<R_2,\text{ and } R_1+R_2<R_0.
\end{align}
To render our energy non-dimensional, we let 
\begin{equation}
\tilde{R}_1=\frac{R_1}{R_0}, \: \tilde{R}_2=\frac{R_2}{R_0}, \:\tilde\gamma=\frac{ \sigma _0R_0}{ \kappa_3 },\: \tilde{\eta}=\frac{\nu R_0^2}{\kappa_3}  \text{ and } \tilde{c} = \dfrac{c}{2\pi ^2R_0^3}.
\end{equation}

Dropping the tilde for better notation and dividing the energy by the constant $2\pi^2 \kappa_3 R_0$, we get
\begin{flalign}
&E\left(R_1,R_2,\phi^{ in},\phi^{ out }\right)=2 \phi^{in} \left(R_2-\sqrt{R_2^2-R_1^2}\right)+ 2\gamma R_1R_2\label{MainEqs}\\
&\quad+\eta\left[\frac{1}{N_1}\phi^{ in} \ln\left(\phi^{ in}\right)+\frac{1}{N_2}\left(1-\phi^{ in}\right) \ln\left(1-\phi^{ in}\right)+\chi \phi^{ in}\left(1-\phi^{ in}\right)\right] R_1^2R_2\nonumber\\
&\quad+\eta\left[\frac{1}{N_1}\phi^{ out } \ln\left(\phi^{ out }\right)+\frac{1}{N_2}\left(1-\phi^{ out }\right) \ln\left(1-\phi^{ out }\right)+\chi\phi^{ out }\left(1-\phi^{ out }\right)\right]\left(\frac{2}{3\pi}- R_1^2R_2\right)\nonumber
\end{flalign}
subject to 
\begin{align}
&\qquad c=\phi^{ in} R_1^2R_2+\phi^{ out }\left[\frac{2}{3\pi}-R_1^2R_2\right],\label{MainConstraints1}\\
&0\leq \phi^{in},\phi^{out} \leq 1,\: 0<R_1<R_2,\text{ and } R_1+R_2<1.
\label{MainConstraints2}
\end{align}

 \subsection{Bending and surface tension effects }
 
 To understand how bending and surface energies compete, and to highlight the contributions of the Flory-Huggins energy, we first consider a reduced model where $\phi^{in}\to 1$ and $\phi^{out}\to 0.$ That is, we assume that all the DSCG material occupies all the space inside the torus. For easier notation, we let $x:= R_1$ and $y := R_2$. We arrive to a simpler minimization problem of the energy
\begin{equation}
E\left(x,y\right)=y-\sqrt{y^2-x^2}+ \gamma xy
\label{ReducedModel}
\end{equation}
subject to 
\begin{align}
&c=x^2y,\:  0<x<y, \label{ReducedConstraints}\\
&\text{ and } x+y<1. \label{ReducedGeomConstraint}
\end{align}
Note that $E(x,y) >0$ for all $x, y$ since $y>\sqrt{y^2-x^2}$ and $\gamma>0$.  \\

We first consider the minimization problem without the geometrical constraint (\ref{ReducedGeomConstraint}).

\begin{theorem}
For any $c,\gamma>0$, $E(x,y)$ subject to the constraints (\ref{ReducedConstraints}) attains a unique non-zero minimum over the interval $(0,\sqrt[3]{c}\;)\times(\sqrt[3]{c}, 1)$. 
\label{ExisUniqTwoConstraints}
\end{theorem}

\begin{remark}
\normalfont The above theorem implies that, ideally, a torus forms for any amount of DSCG added to the bath and any elastic coefficients considered.
\end{remark}
\begin{proof}
We replace $y$ from (\ref{ReducedConstraints}a) in (\ref{ReducedModel}), assuming that $x\neq 0$. Then,
\begin{align}
E(x) & = \frac{c}{x^2}-\frac{\sqrt{c^2-x^6}}{x^2}+\frac{\gamma c}{x}.&
\label{ReducedModOneVar}
\end{align}
To find the critical points, we need to study $E'(x)=0.$
\begin{align*}
E'(x) & = \frac{1}{x^3}\left[-2c-\gamma c x+\frac{x^6+2c^2}{\sqrt{c^2-x^6}}\right]. 
\end{align*}
To solve $E'(x)=0$, it is sufficient to solve
\[\frac{\left(x^3/c\right)^2+2}{\sqrt{1-\left(x^3/c\right)^2}} = 2+\gamma  x.\]
We let $s:=\frac{x^3}{c}$, with the condition that $0< s<1$ (from \ref{ReducedConstraints}b). Finding critical points now reduces to finding the point(s) of intersection, if any, of the two curves
\[f(s) =\frac{s^2+2}{\sqrt{1-s^2}}\qquad \text{and } \qquad g(s)= 2 + \gamma \sqrt[3]{c}\sqrt[3]{s},\]
over the interval $[0,1).$ Note that
\[f'(s)= \frac{s(2-s)(2+s)}{2(1-s^2)},\qquad f''(s)=\frac{2(s^4+s^2+4)}{4(1-s^2)^2},\]
\[g'(s)= \frac{\gamma \sqrt[3]{c}}{3(\sqrt[3]{s})^2}, \qquad g''(s)=-\frac{2\gamma\sqrt[3]{c}}{9(\sqrt[3]{s})^5}.\]
The function $f$ has the following properties:
\begin{list}{$\circ$}{}  
\item $f(0) = 2$
\item $\displaystyle\lim_{s\to 1}f(s) = +\infty$
\item $f'(s)>0$ for $0<s<1$, so $f$ is increasing.
\item $f''(s)>0$ so $f$ is concave up. 
\item $f'(0)=0$
\end{list}
and the function $g$ has the following properties:
\begin{list}{$\circ$}{}  
\item $g(0) = 2$
\item $g(1) = 2+ \gamma \sqrt[3]{c}$
\item $g'(s)>0$ for $s\neq 0$, so $g$ is increasing.
\item $g''(s)<0$ for $0<s<1$ so $g$ is concave down. 
\item $\displaystyle\lim_{s\to 0}g'(s)=+\infty$
\end{list}

Both functions are continuous over (0,1), $f$ has a horizontal tangent line at $(0,2)$, while $g$ has a vertical tangent line at that point. Both are increasing over that interval, with opposite concavities, and $f$ has a vertical asymptote at $s=1.$ This implies that, for any $c,\gamma >0$, there is always an intersection point over the interval $(0,1)$, other than at $s=0$. Hence, $E'(x)=0$ has one solution over the interval $(0,\sqrt[3]{c}),$ for any $c,\gamma >0$. Since $y = c/x^2$, then $y\in(\sqrt[3]{c}, 1)$. To determine if the critical point is indeed a minimum, we calculate the second derivative.
\begin{align*}
E''(x) & = \frac{6c}{x^4}+\frac{2\gamma c}{x^3}+\frac{3x^5[3c^2+(c^2-x^6)]}{(c^2-x^6)\sqrt{c^2-x^6}}.&
\end{align*}
We have $c,\gamma,x>0$ and since $x^3<c$, then $c^2-x^6>0$. So $E''(x)>0$ for all $x>0.$ In particular, $E''(x_1)>0$, where $x_1$ is a non-zero critical point. Hence, any non-zero critical point is a minimum.
\end{proof}

\begin{figure}[htb!]
    \centering
    \begin{minipage}{1.0\textwidth}
        \centering
        \includegraphics[scale=0.35]{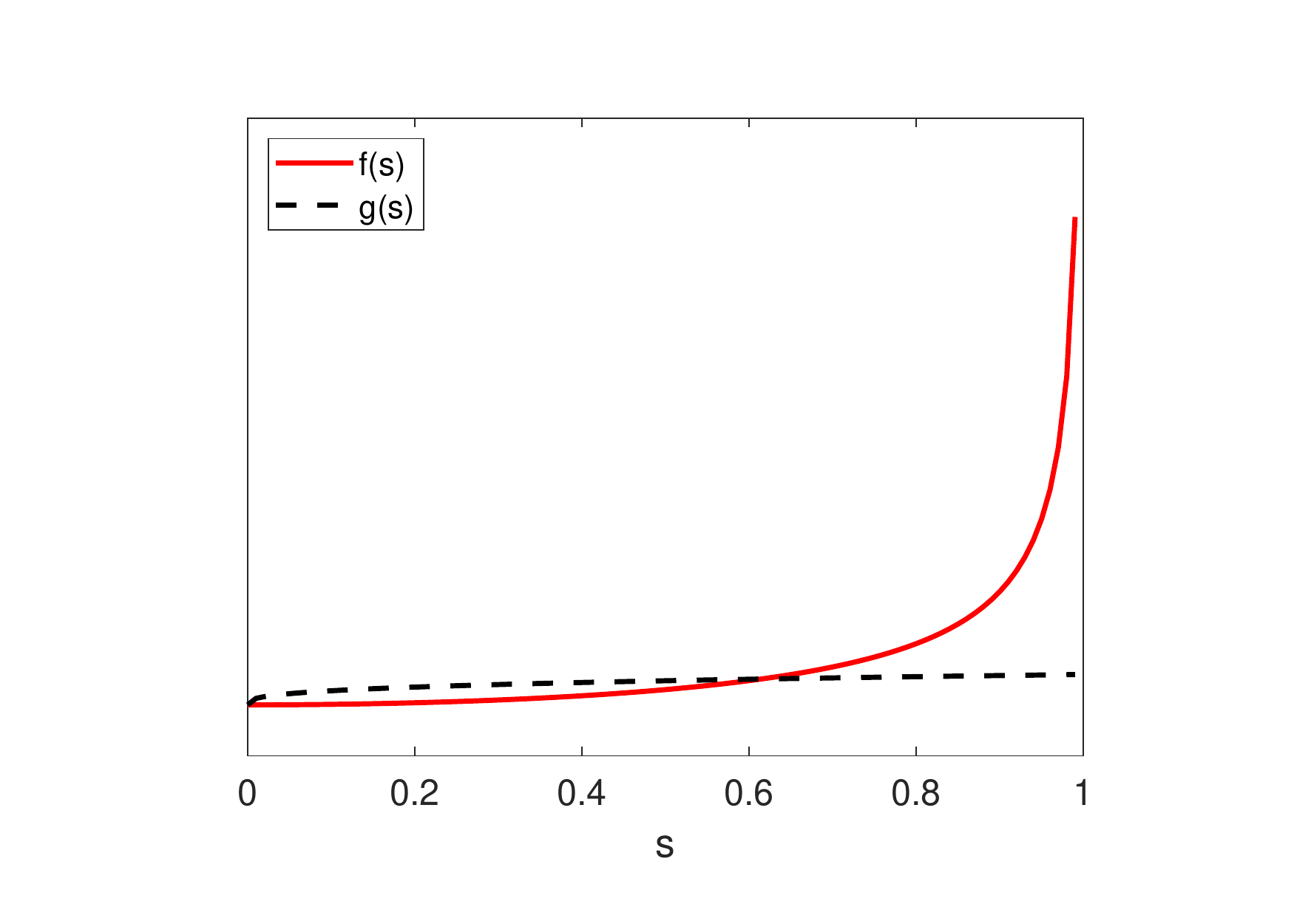}\includegraphics[scale=0.35]{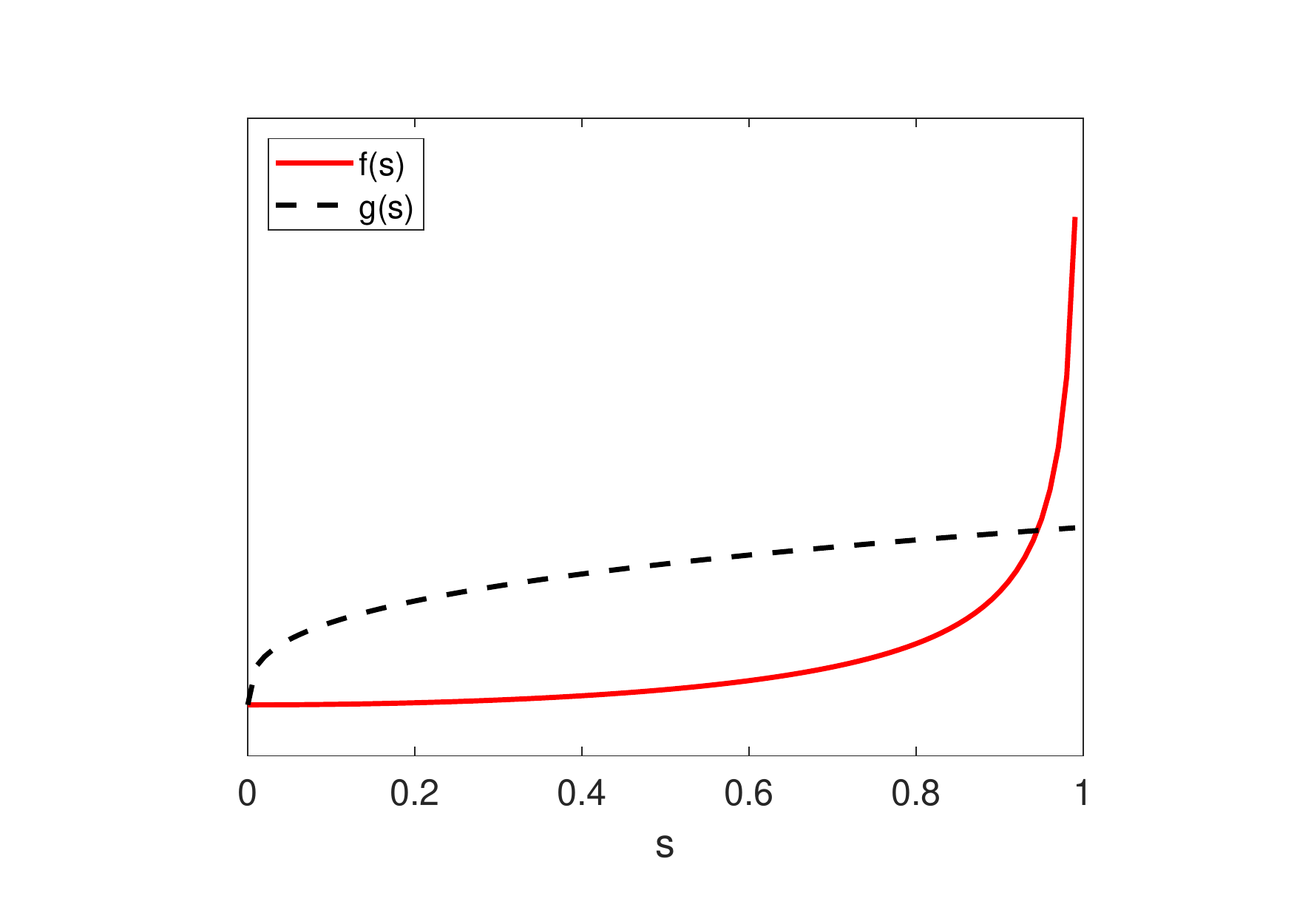}\\
    \end{minipage}
    \caption{Intersection of $f(s)$ and $g(s)$ for $c=0.0005$ (left) and $c=0.1$ (right), and $\gamma=15$.}
    \label{figure:Int_solutions}
\end{figure}

Since the proof of Theorem \ref{ExisUniqTwoConstraints} was constructive, we can numerically calculate the minimum of (\ref{ReducedModel}) subject to (\ref{ReducedConstraints}) (see Figure~\ref{figure:Int_solutions}). We can then deduce the values of $(x,y)$ for different values of $c$ and $\gamma.$ 

To solve the minimization problem (\ref{ReducedModel}) subject to (\ref{ReducedConstraints}) and (\ref{ReducedGeomConstraint}), we still need to enforce the geometric constraint (\ref{ReducedGeomConstraint}): $x+y<1$. If we replace $y$ by $c/x^2$, we get the inequality $x^3-x^2+c<0.$ We prove the following result about this cubic inequality.

\begin{lemma}\label{cubicSol}
If $0<c<4/27$, then $x^3-x^2+c<0$ when $x_2<x<x_3$, where $x_2$ and $x_3$ are the two positive roots of $x^3-x^2+c=0.$
\end{lemma}

\begin{figure}[htb!]
\centering
\includegraphics[width=0.6\textwidth]{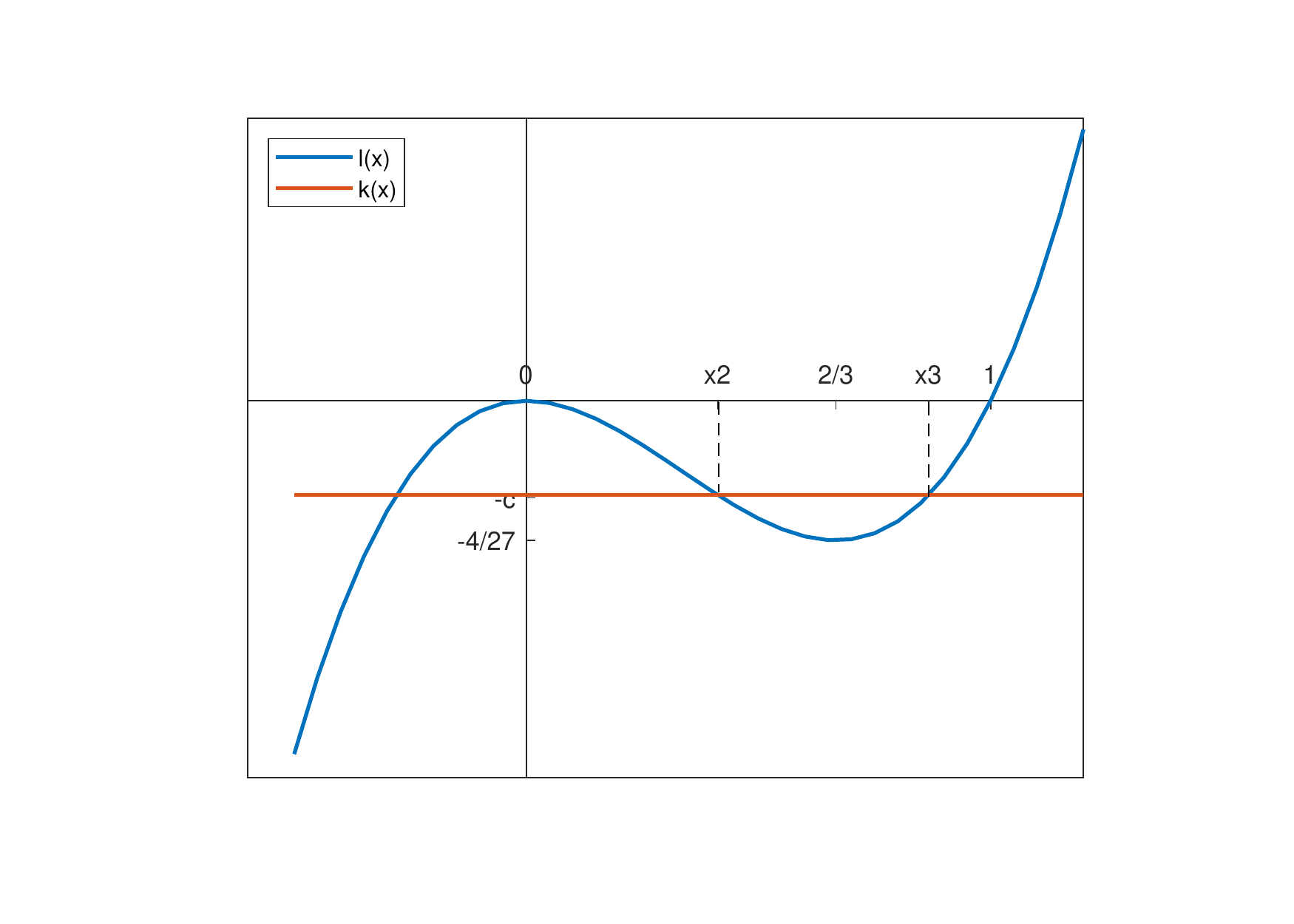}
\caption{Enforcing the geometric constraint (\ref{ReducedGeomConstraint}): $x^3-x^2+c<0$.}
\label{figure:cubic_Inequality}
\end{figure}

\begin{proof}
Rewrite the inequality $x^3-x^2+c<0$ as $l(x)<k(x)$, where $l(x) = x^3-x^2$ and $k(x) = -c.$ The cubic function $l(x)$ has a local minimum at $2/3$ with $l(2/3)= -4/27.$ Assume $0<c<4/27$ and let $x_2$ and $x_3$ be the positive points of intersection of the cubic function with the horizontal line $k(x) = -c$. Note that $x_2$ and $x_3$ are also the positive roots of $x^3-x^2+c=0$. Then, $x^3-x^2+c$ is negative whenever  $x_2<x<x_3$. (See Figure~\ref{figure:cubic_Inequality}).
\end{proof}

The following minimization result then follows from Lemma \ref{cubicSol} and Theorem \ref{ExisUniqTwoConstraints}.
\begin{corollary}
Suppose $(x, c/x^2)$ minimizes (\ref{ReducedModel}) subject to (\ref{ReducedConstraints}). If $x_2<x<\min(x_3, \sqrt[3]{c})$, then $(x, c/x^2)$ is the unique minimum of (\ref{ReducedModel}) subject to (\ref{ReducedConstraints}) and (\ref{ReducedGeomConstraint}).
\label{ExisUniq}
\end{corollary}

\subsection{Experimental setup and physical constants}

\begin{figure}[htb!]
\centering
\includegraphics[width=0.7\textwidth]{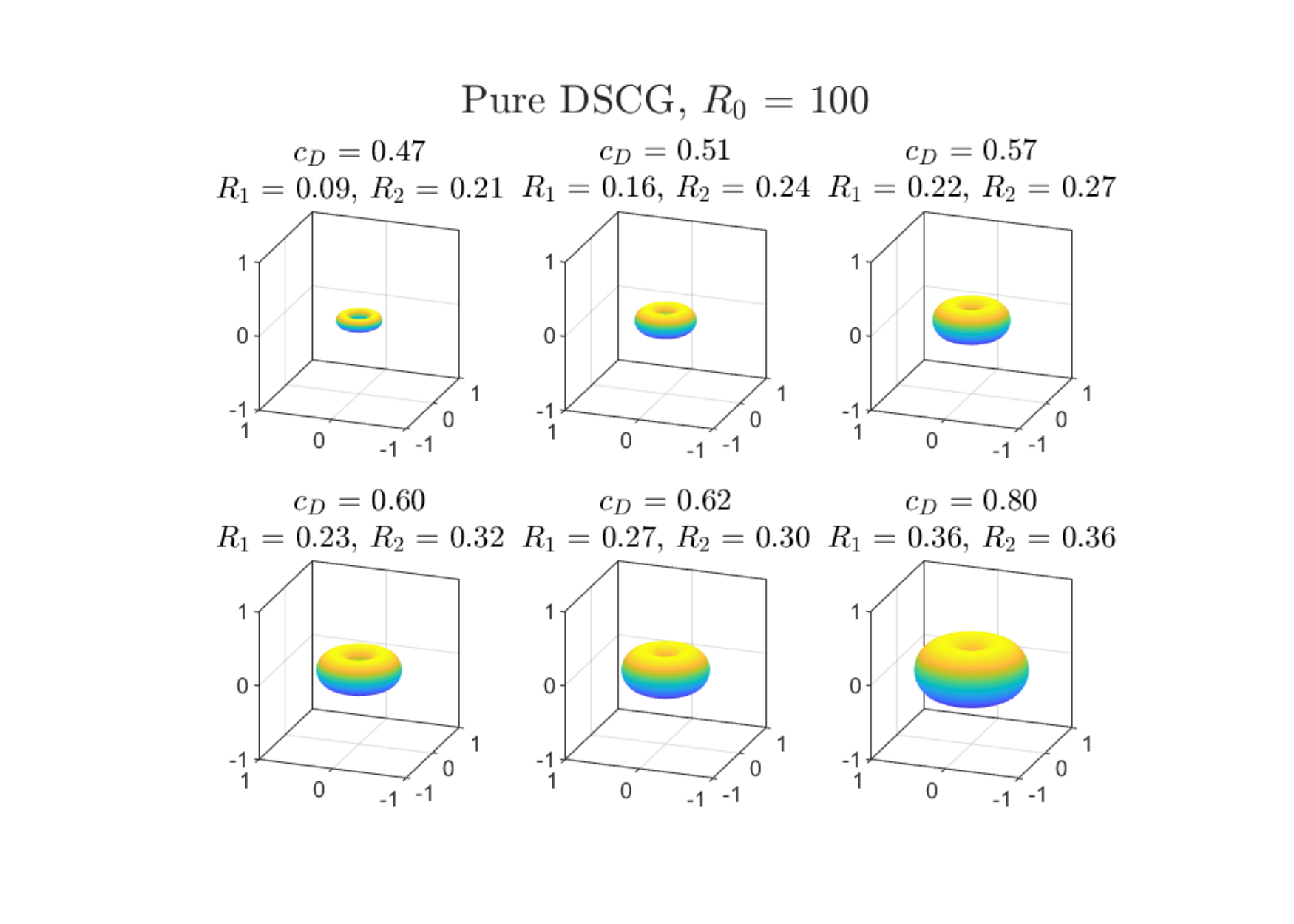}
\caption{Non-dimensionalized average experimental data with pure DSCG.}
\label{fig:ExperTori}
\end{figure}

We turn now to applying our minimization result in Corollary~\ref{ExisUniq} in order to simulate the experimental data we estimated from \cite{koizumi2022}. We manually extracted the data from relevant figures using grabit, a Matlab GUI \cite{grabit}. The experiments were performed in rectangular glass capillaries of width 4 mm and thickness 0.2mm. The mixture of DSCG in water is cooled down from the isotropic phase, and within 20 minutes after cooling, toroidal structures tend to stabilize. Measurements reported in \cite{koizumi2019} identify this phenomenon with the I+C phase separation, that is, nucleation of columnar hexagonal structures on their own isotropic liquid. Samples with different concentrations of DSCG were tested in six experiments in this case. The values of the concentrations are shown in Table \ref{tab:ExperData}, along with the average inner and outer radii observed. To compare our results to the experimental observations, we non-dimensionalize the values observed by assuming a typical length $R_0 = 100\mu$m. Figure~\ref{fig:ExperTori} visualizes the tori with their non-dimensionalized radii.

\begin{table}[h]
\caption{Average experimental data for pure DSCG, obtained from figure 1(b) in \cite{koizumi2022}.}
\label{tab:ExperData}
\centering
\begin{tabular}{cccc} \hline 
Exper& $\cd$ & $R_1$ & $R_2$  \\
& (in mol/Kg)& (in $\mu$m) & (in $\mu$m) \\ \hline\hline
1 & 0.47 & 8.68 & 20.75\\ \hline
2 & 0.51 & 15.70 &  23.61\\  \hline
3 & 0.57 & 22.12 &  27.49\\  \hline
4 & 0.6 & 22.61 & 31.79\\  \hline
5 & 0.62 & 26.58 & 29.61\\  \hline
6 & 0.8 & 36.40 & 36.40\\  \hline
\end{tabular}
\end{table}

We note that the authors in \cite{koizumi2022} consider a numerical model that captures the faceted shapes of the toroids and half-toroids for the same set of experiments. The difference in our work, other than considering different terms of the energy density and emphasizing a different set of parameters, is that we do not seek a topological description of the observations but rather we aim to identify how the liquid crystal order competes with the aggregation phenomena.

We now turn to calculate the energy ratio of surface tension to bending, non-dimensionalized as the $\gamma$ parameter in our notation, in Table \ref{tab:ExperGamma} from a relationship between $R_1$, $R_2$, and $\gamma$ that can be obtained from $E'(R_1) = 0$, namely,
\begin{equation}\label{eq: gamma}
\frac{\left(R_1/R_2\right)^2+2}{\sqrt{1-\left(R_1/R_2\right)^2}} = 2+\gamma  R_1.
\end{equation}

\begin{table}[htb!]
\caption{Deduced $\gamma$ values from average experimental data.}
\label{tab:ExperGamma}
\centering
\begin{tabular}{c||@{\hskip 3pt}c@{\hskip 10pt}c@{\hskip 10pt}c@{\hskip 10pt}c@{\hskip 10pt}c@{\hskip 10pt}c} \hline 
Experiment& 1& 2& 3& 4& 5& 6\\ \hline
 $\gamma$ & 4.5456 & 8.0886 & 11.1165 & 6.9204 &  16.4303 & \textit{large} \\\hline
\end{tabular}
\end{table}

\begin{figure}[!hb]
\centering
\includegraphics[width=0.7\textwidth]{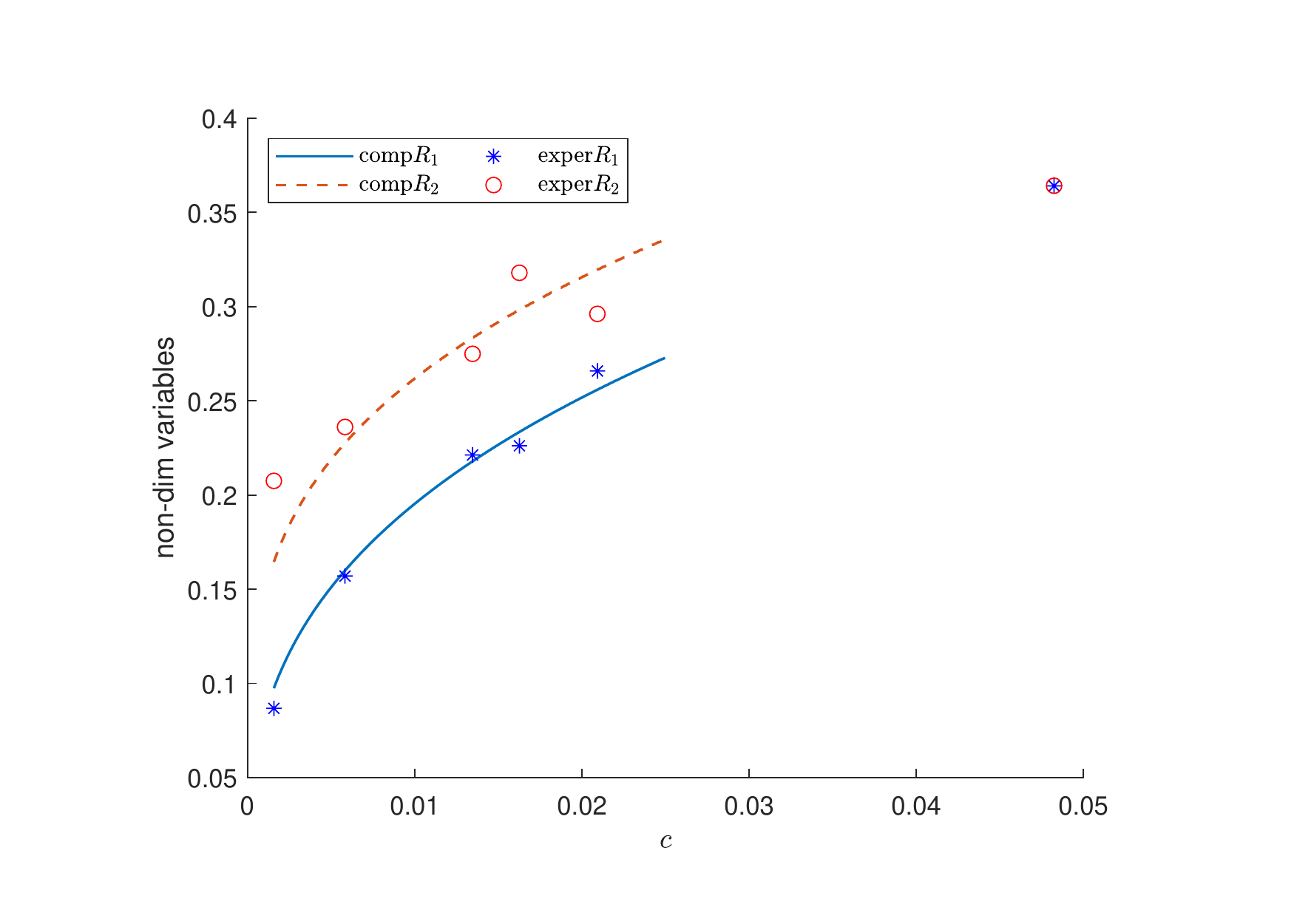}
\caption{Experimental versus computed data with an average $\gamma$ approximation, when taking into account only bending and surface tension effects. Since the value of $\gamma$ from the last experiment is too large compared to all other values, it is excluded from the computations.}
\label{fig:ExpervsComp}
\end{figure}

We discuss this energy ratio further in Section~\ref{sec: gamma} but for the sake of these simulations, we simply take the average of $\gamma$ from the first five experiments, $\gamma_{avg}= 9.42$. Note that the last value is large because $R_1 \approx R_2$, so we exclude it from the average. For the parameter $c$, we use the values of exper$R_1^2$exper$R_2$. Applying the methods outlined in Theorem \ref{ExisUniqTwoConstraints} and Corollary \ref{ExisUniq}, we then calculate the values of comp$R_1$ and comp$R_2$. The results are in Table~\ref{tab:ExpervsComp} and Figure~\ref{fig:ExpervsComp}. Note that the errors are smaller when the value of $\gamma$ is closer to the experimental value.

\begin{table}[h!]
\caption{Average experimental versus computed non-dimensional data for pure DSCG experiments.}
\label{tab:ExpervsComp}
\centering
\begin{tabular}{c@{\hskip 5pt}ccc@{\hskip 10pt}cc@{\hskip 10pt}cc} \hline 
Exper& exper$R_1$ & exper$R_2$ & exper$Vol$ & comp$R_1$ & comp$R_2$& error$R_1$ & error$R_2$\\ \hline\hline
1 & 0.0868 & 0.2075& 0.0016 & 0.0983 & 0.1654& 0.0115 & 0.0421\\ \hline
2 & 0.1570 & 0.2361& 0.0058 & 0.1597 & 0.2274& 0.0027 & 0.0087\\  \hline
3 & 0.2212 & 0.2749& 0.0135 & 0.2182 & 0.2836& 0.0030 & 0.0087\\  \hline
4 & 0.2261 & 0.3179 & 0.0163& 0.2337 & 0.2983& 0.0076 & 0.0196\\  \hline
5 & 0.2658 & 0.2961 & 0.0209& 0.2558 & 0.3195& 0.0100& 0.0234\\  \hline
6 & 0.3640 &  0.3641 & 0.0482 & - & -& - & - \\  \hline
\end{tabular}
\end{table}

\subsection{Values of surface to bending energy ratio}\label{sec: gamma}
It is evident that the calculated value of $\tilde\gamma$ affects the quality of the simulated results, so we explore this further and compare our results to experimental data. Replacing $\tilde\gamma$ by its definition $\sigma_0R_0/\kappa_3$ in (\ref{eq: gamma}),  leads to

\begin{equation}
\frac{\sigma_0}{\kappa_3} = \frac{\left(R_1/R_2\right)^2+2}{R_1\sqrt{1-\left(R_1/R_2\right)^2}}-\frac{2}{R_1}.
\label{KSigma}
\end{equation}
Note that $R_1$ and $R_2$ here are the lengths of the radii in $\mu$m. Thus, Equation~\eqref{KSigma} allows us to calculate $\sigma_0/\kappa_3$ directly from the tori dimensions. If the ratio $R_1/R_2$ is small, the ratio can be estimated by 
\begin{equation}
\frac{\sigma_0}{\kappa_3} = \frac{2+2\left(R_1/R_2\right)^2}{R_1}-\frac{2}{R_1} = 2\frac{R_1}{R_2^2}.
\label{KSigma_Est}
\end{equation}
Tortora \textit{et. al.} \cite{lavrentovich2010} give a formula to calculate $\sigma_0/\kappa_3$. Theirs is a formula based on minimizing bending and surface energy densities subject to a volume constraint. The difference in our work is that we have an exact formula for the bending energy, whereas they approximate it by $\kappa_3V/R_2^3$. Their expression for a torus is
\[\frac{R_1}{R_2} = \frac{1}{(2 \pi^2)^{1/5}}\left(\frac{\sigma_0}{\kappa_3}\right)^{3/5}V^{1/5} = \frac{1}{(2 \pi^2)^{1/5} \beta^{3/5}},\]
where $\beta = \dfrac{\kappa_3}{\sigma_0 V^{1/3}}$ is a dimensionless parameter.

Table \ref{tab:ExperData_ksigma} shows the experimental values of $\sigma_0/\kappa_3$ for experiments with pure DSCG \cite{koizumi2022}. Note that the values of $\beta$ given in figure 3(b) therein correspond to measurements taken for half tori. Since we are considering full tori, we calculate our value of $\beta$ using the formula $\beta_{\text{full}} = \beta_{\text{half}}/2^{1/3}$ since $V_{\text{full}} = 2 V_{\text{half}}$. We also remark that the volume in the experiments is calculated using the formula $\pi R_2A$, where $A$ is found experimentally. Table \ref{tab:ksigma} compares the experimental values of $\sigma_0/\kappa_3$ to the exact and estimated values using Equations~\eqref{KSigma} and \eqref{KSigma_Est}, respectively.

\begin{table}[htb!]
\caption{Average experimental data of $\sigma_0/\kappa_3$ in case of pure DSCG.}
\label{tab:ExperData_ksigma}
\centering
\begin{tabular}{c@{\hskip 5pt}c@{\hskip 15pt}c@{\hskip 15pt}c@{\hskip 10pt}c} \hline 
Exper& $\cd$ & $V$ & $\beta$ & $\sigma_0/\kappa_3=1/(\beta V^{1/3})$  \\
& (in mol/Kg)& (in $\mu$m$^3$) & & (in 1/$\mu$m) \\ \hline\hline
1 & 0.47 & 49166 & 1.3250
 & 0.02060\\ \hline
2 & 0.51 & 231330 &  0.7806
& 0.02087\\  \hline
3 & 0.57 & 503517 &  0.5951
& 0.02112\\  \hline
4 & 0.6 & 985269 & 0.4671
& 0.02152\\  \hline
5 & 0.62 & 1394959 & 0.4162
& 0.02150\\  \hline
6 & 0.8 & 2474051 & 0.3369
 & 0.02195\\  \hline
\end{tabular}
\end{table}

\begin{table}[hb!]
\caption{Experimental versus calculated values of $\sigma_0/\kappa_3$.}
\label{tab:ksigma}
\centering
\begin{tabular}{ccccc} \hline 
Exper& $\cd$ & exper($\sigma_0/\kappa_3$) & $\sigma_0/\kappa_3$ & est($\sigma_0/\kappa_3$) \\
& (in mol/Kg)& (in 1/$\mu$m) & (in 1/$\mu$m) & (in 1/$\mu$m) \\ \hline\hline
1 & 0.47 & 0.02060 & 0.0455& 0.0403\\ \hline
2 & 0.51 & 0.02087 & 0.0809&0.0563\\  \hline
3 & 0.57 &  0.02112 & 0.1112& 0.0585\\  \hline
4 & 0.6 &  0.02152 & 0.0692& 0.0447\\  \hline
5 & 0.62 & 0.02150 & 0.1643& 0.0606\\  \hline
6 & 0.8 &  0.02195 & \textit{large} & 0.0549\\  \hline
\end{tabular}
\end{table}

 \subsection{Cylinder rearranged as a torus}\label{CylindertoTorus} 
We conclude this section estimating the  threshold value of the surface energy on the circular bases of a cylinder of volume $V_0$, with the director field arranged parallel to the axis, so that it will adopt a torus form in order to reduce energy. This is the result of the hydrophobic nature of the broad face of the disk-like molecules.  A simple calculation shows that the total energy of a cylinder of height $H=2\pi R_2$ is
\begin{align}
    E_{\text{\begin{small}{cyl}\end{small}}} = 4\pi^2 \sigma_0R_1R_2 + 2\sigma_1 \pi R_1^2 
    = 2\sigma_0 \frac{V_0}{R_1}+ 2\sigma_1\pi R_1^2,
\end{align}
where $\sigma_1$ is the surface tension coefficient of the circular base of the cylinder. This energy equates that of a torus with radius $R_1$ and $R_2$, with prescribed volume $V_0$ as in the previous section, when
\begin{equation}
    \frac{\sigma_1}{\kappa_3} R_1^2=2\pi\left(\frac{V_0}{2\pi^2 R_1^2} -\sqrt {\frac{V_0^2}{4\pi^4 R_1^4}-R_1^2}\right).
\end{equation}
Moreover, $R_1$ in the previous relation should be taken as the optimal value previously computed.
The above relation gives an estimate of the value of $\sigma_1$ needed for a cylinder of volume $V_0$, enveloping a family of straight vector lines,  to rearrange itself  into a torus.

\section{Flory-Huggins energy: phase separating regime}\label{sec:FH}
  
  In this section, we are concerned with the  Flory-Huggins energy in the phase separating regime. Note that the logarithmic terms in the function combine to produce a minimum  for values of $\phi$ well inside the interval $(0,1)$, so such terms contribute to the mixing of both species.  However, the $\chi$ term is positive in the interior of the interval and vanishes at  the ends, $\phi=0, 1$. That is, such a term is representative of the repulsive mechanism that tends to keep the phase separated. Hence, $\chi$ has a primary role in controlling the convexity of the Flory-Huggins energy function that occurs, as will be shown below.
     
      Recall the nondimensional Flory-Huggins energy density in (\ref{FH_energy_2}), 
\[E_{\mbox{FH}} = \nu \int_{\mathcal{B}}H(\phi)d\mathbf{x},\]      
where
 \[H(\phi) := \frac{1}{N_1}\phi  \ln\left(\phi\right)+\frac{1}{N_2}\left(1-\phi \right) \ln\left(1-\phi\right)+\chi \phi\left(1-\phi\right).\]
 Phase separation into two phases occurs if the stability criterion $H''(\phi)>0$ is not satisfied for some values of $\phi.$ At the critical value of Flory parameter, $\chi_c,$ the second derivative of $H(\phi)$ is positive everywhere except at the critical volume fraction $\phi_c$ where it is zero~\cite{Brinke}. So the critical point corresponds to the minimum of $H''(\phi)$ and is determined by
\[H''(\phi) = H'''(\phi) = 0.\]

Therefore, $\phi_c$ and $\chi_c$ satisfy the system 
\[\begin{cases}
H''(\phi)= \dfrac{1}{N_1\phi}+\dfrac{1}{N_2(1-\phi)}-2\chi = 0,\\[2ex]
H'''(\phi)=-\dfrac{1}{N_1\phi^2}+\dfrac{1}{N_2(1-\phi)^2} = 0.
\end{cases}\]

Solving the above system, we get
\[\phi_c = \frac{\sqrt{N_2}}{\sqrt{N_1}+\sqrt{N_2}}, \quad \chi_c = \frac{1}{2}\left[\frac{1}{\sqrt{N_1}}+\frac{1}{\sqrt{N_2}}\right]^2.\]

We can prove that phase separation occurs when $\chi > \chi_c.$ Assume $\chi > \chi_c$ and calculate $H''(\phi_c).$
\begin{flalign*}
H''(\phi_c) & = \frac{1}{N_1\phi_c}+\frac{1}{N_2(1-\phi_c)}-2\chi =  2\chi_c - 2 \chi  = -2 (\chi - \chi_c),
\end{flalign*}
 which is negative since $\chi > \chi_c$. Away from $0$ and $1,$ $H''(\phi)$ is a continuous function so there exists an interval $I_c$ containing $\phi_c$ over which $H''$ is negative. 

\begin{table}[h!]
\caption{Values of $\chi_c$, $\phi_c$, and $\ca$ for different values of $N_1$.}
\label{tab:chi,phi}
\centering
\begin{tabular}{ccccc} 
\hline
$N_1$ & $N_2$ & $\chi_c$ & $\phi_c$& ${\ca}$\\
\hline\hline
1 & 1 & 2& 0.5& 1.5\\ 
\hline
50 & 1& 0.6514 & 0.1239 & 0.641\\ 
\hline
100 & 1& 0.6050 & 0.0909& 0.6\\
\hline
250 & 1& 0.5652 & 0.0595& 0.5632\\
\hline
\end{tabular}
\end{table}

\begin{figure}[h!]
    \centering
  \includegraphics[scale=0.4]{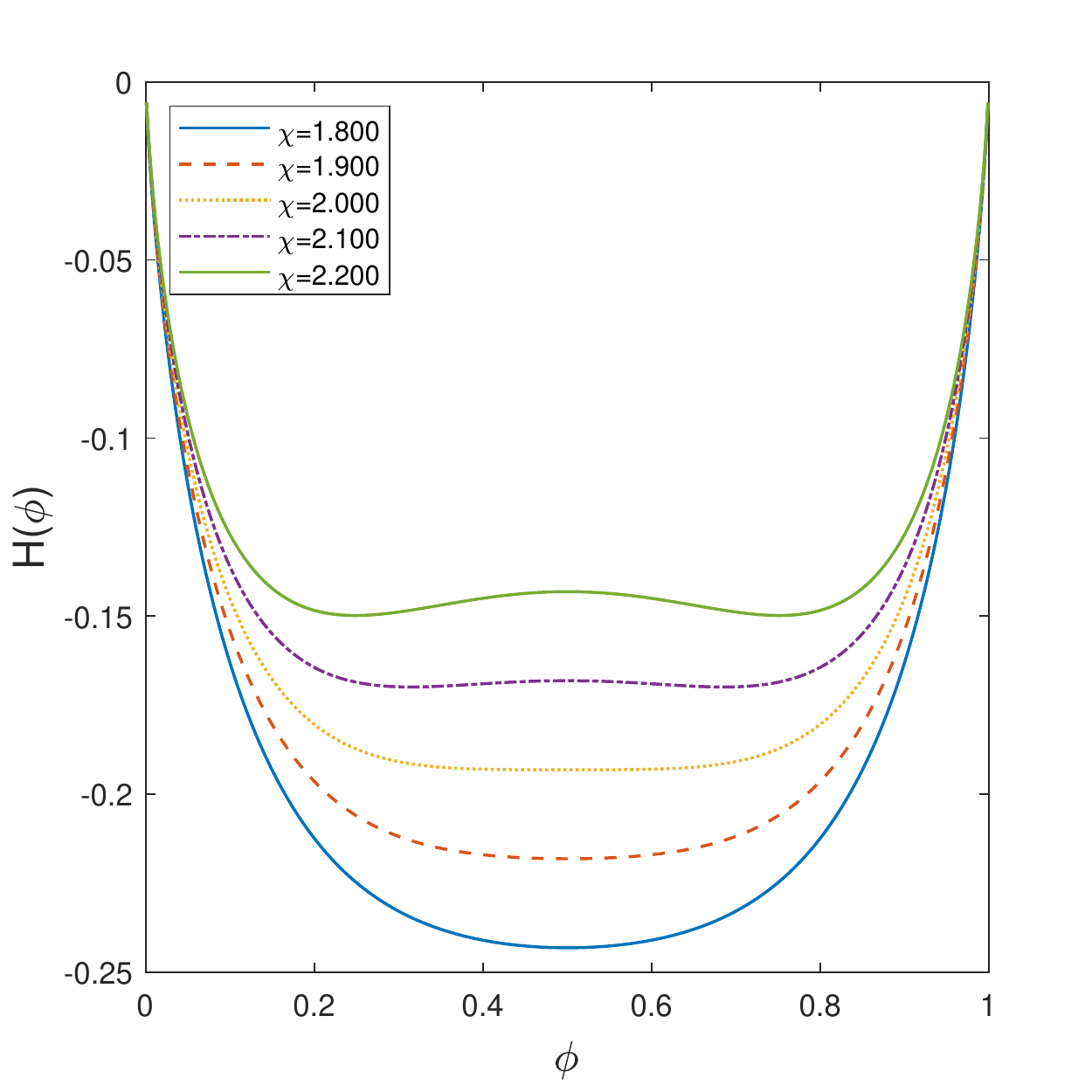}\hspace{0.2in}
   \includegraphics[scale=0.4]{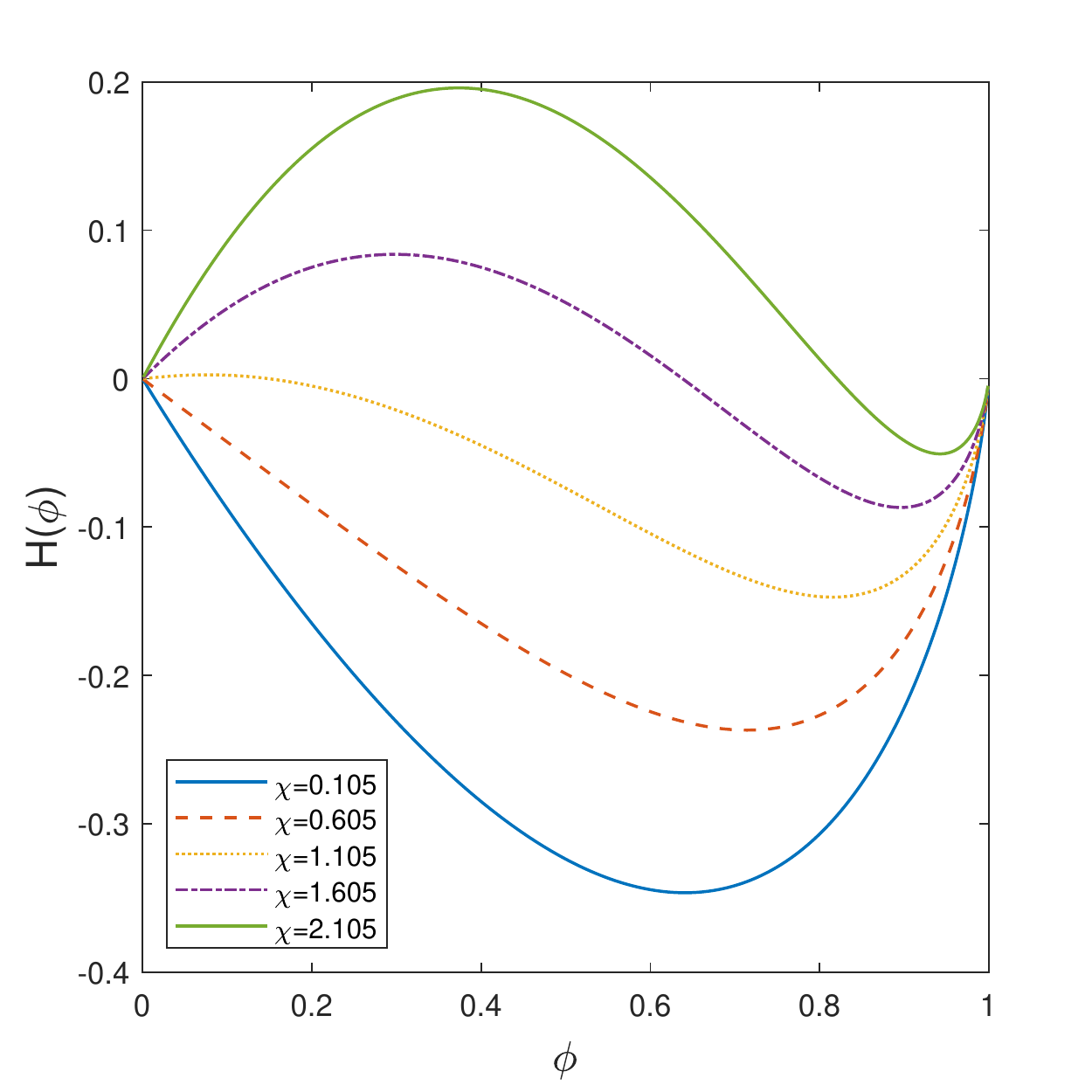}
   \caption{ Flory-Huggins energy density plots for $N_1=1,\: N_2=1, \: \chi_c = 2$ along with several values close to $\chi_c$ (left) and for $N_1=100,\: N_2=1,\: \chi_c = 0.605$ along with several values close to $\chi_c$ (right).}
    \label{fig:FH_N1_1_100}
\end{figure}

Table~\ref{tab:chi,phi} lists some critical values of $\chi$ and $\phi$ corresponding to specific values of $N_1$. The graph of the Flory-Huggins energy density is shown in Figure~\ref{fig:FH_N1_1_100} with values corresponding to $\chi_c$ and several other values lower and higher than it, for the cases $N_1=N_2=1$ and $N_1=100, N_2=1$.

We are interested in the case where $N_2 =1$ and $N_1 = N >>1.$ We approximate the critical value of $\chi$, $\chi_c = \frac{1}{2}\left[\frac{1}{\sqrt{N}}+1\right]^2= \frac{1}{2}+\frac{1}{\sqrt{N}}+\frac{1}{2N},$ to get
\[\ca =  \frac{1}{2}+\frac{1}{\sqrt{N}},\]
in agreement with the value found by de Gennes \cite{de1979scaling}. The last column of Table \ref{tab:chi,phi} shows that this approximate value, $\ca$, is a good estimate starting from $N_1=100$. 

\subsection{Full model simulation including the condensing agent PEG}  
\begin{figure}[!ht]
\centering
\includegraphics[width=0.7\textwidth]{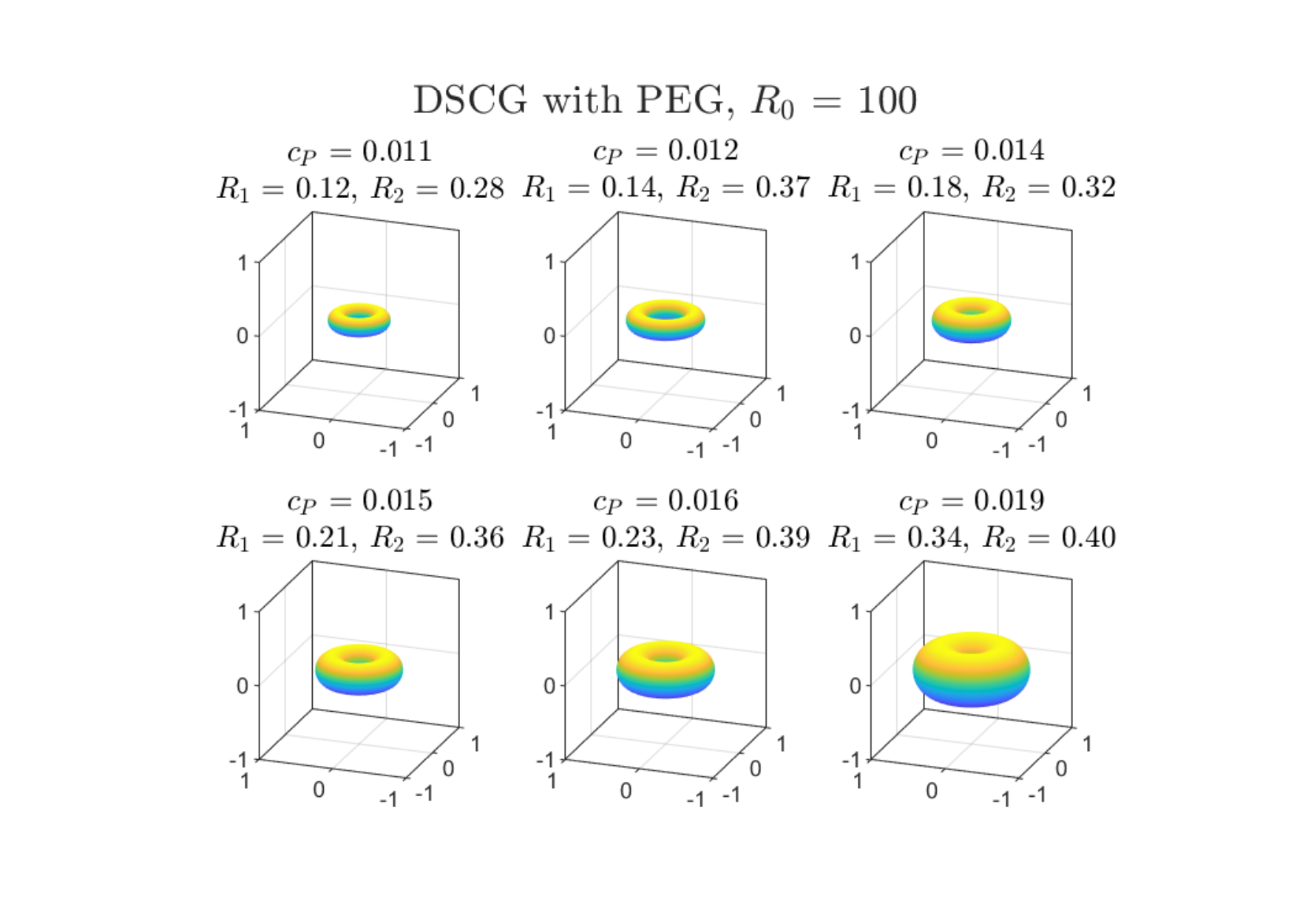}
\caption{Non-dimensionalized average experimental data of DSCG with PEG.}
\label{fig:ExperTori_PEG}
\end{figure}

Table \ref{tab:ExperData_PEG} contains the experimental data of six experiments with PEG \cite{koizumi2022}. The DSCG concentration is fixed at 0.34 mol/Kg. For each concentration of PEG, the inner and outer radii are given in $\mu$m. Figure~\ref{fig:ExperTori_PEG} visualizes the tori with their non-dimensionalized radii.
\begin{table}[htb!]
\caption{Average experimental data for DSCG with PEG, obtained from figure 1(g) in \cite{koizumi2022}.}
\label{tab:ExperData_PEG}
\centering
\begin{tabular}{cccc} \hline 
Exper& $\cp$ & $R_1$ & $R_2$  \\
& (in mol/Kg)& (in $\mu$m) & (in $\mu$m) \\ \hline\hline
1 & 0.011 & 12.3 & 27.9 \\ \hline
2 & 0.012 & 13.67 &  36.78 \\  \hline
3 & 0.014 & 18.29 &  32.3 \\  \hline
4 & 0.015 & 20.55 & 35.53 \\  \hline
5 & 0.016 & 23.29 & 39.33 \\  \hline
6 & 0.019 & 34.15 & 40.47 \\  \hline
\end{tabular}
\end{table}

\bigskip

To find numerical solutions to the full model (\ref{MainEqs}) subject to (\ref{MainConstraints1}) and (\ref{MainConstraints2}), we will use the Particle Swarm Optimization (PSO) method \cite{PSO}. This method solves a minimization problem, taking into account its constraints, by evaluating a population of candidate solutions (``particles", i.e., points in the $n$-dimensional space of variables) and updating the particle positions iteratively according to a simple formula. Each particle's movement is influenced by its local best known position (i.e., the one yielding the minimum function value so far), but is also guided toward the global best position of the swarm, which is updated as better positions are found by other particles. This moves the swarm toward the best solution to the minimization problem. Although conceptually simple, such a population-based approach offers several advantages over gradient-based methods, including not requiring derivative information, ease of handling of constraints, and a higher probability of converging to the global minimum due to the spread of the swarm.

Recall that the four variables for the full model are $R_1,\: R_2,\: \phi^{in},$ and $\phi^{out}.$ We fix the following parameters
\[\gamma = 2.8956,\quad N_1 = 100, \quad N_2 = 1, \quad \eta = 1000.\]
Note that $\gamma$ value is approximated from the experimental values, with PEG, as in Section~\ref{sec: gamma}. We consider values of $\chi$ greater than the critical value $1/2+1/\sqrt{N_1}$, and then find numerical solutions over an interval of the parameter $c$. The goal is to find the value of $\chi$ that captures the experiments with PEG most closely.

\begin{figure}[!ht]
\centering
\includegraphics[width=0.45\textwidth]{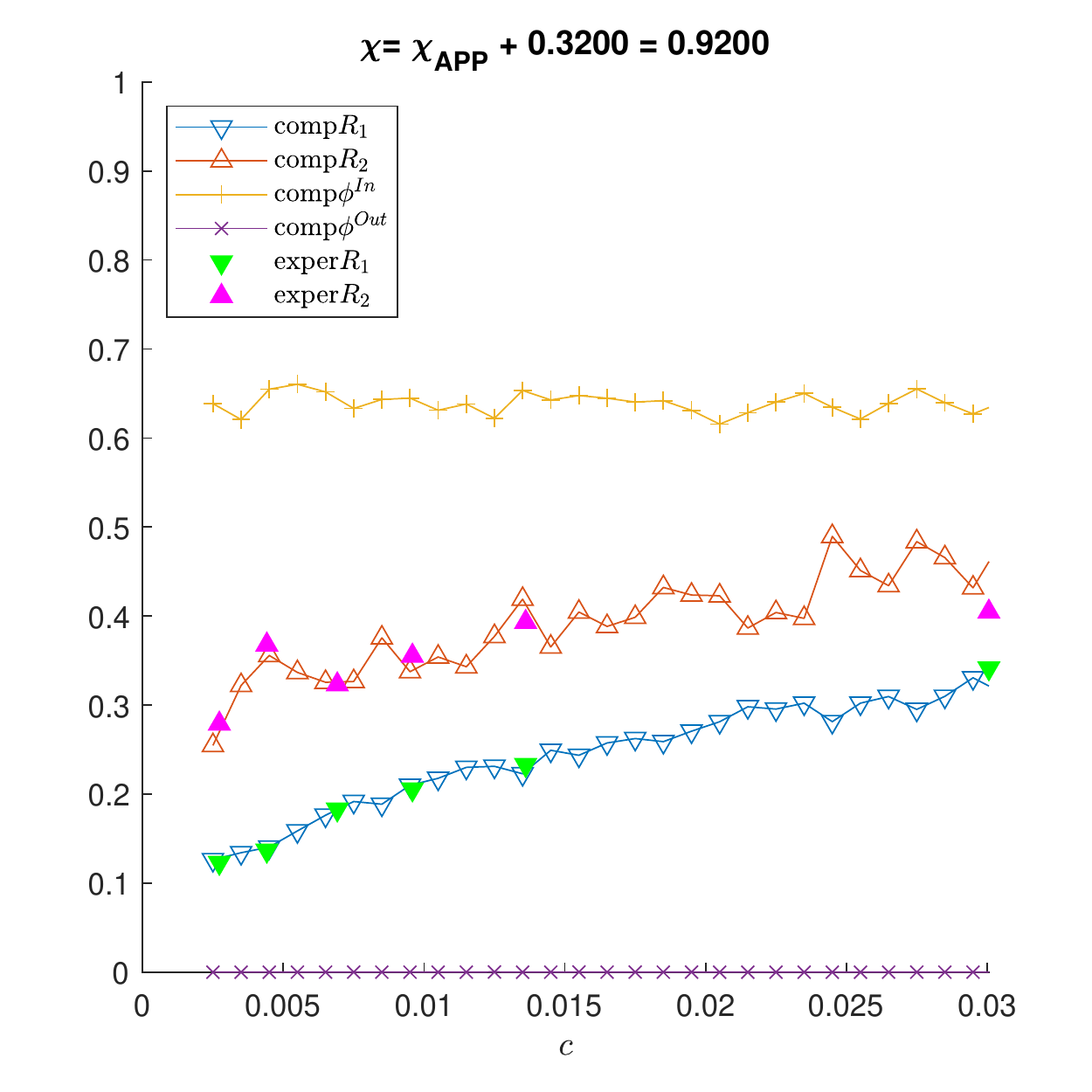}
\includegraphics[width=0.45\textwidth]{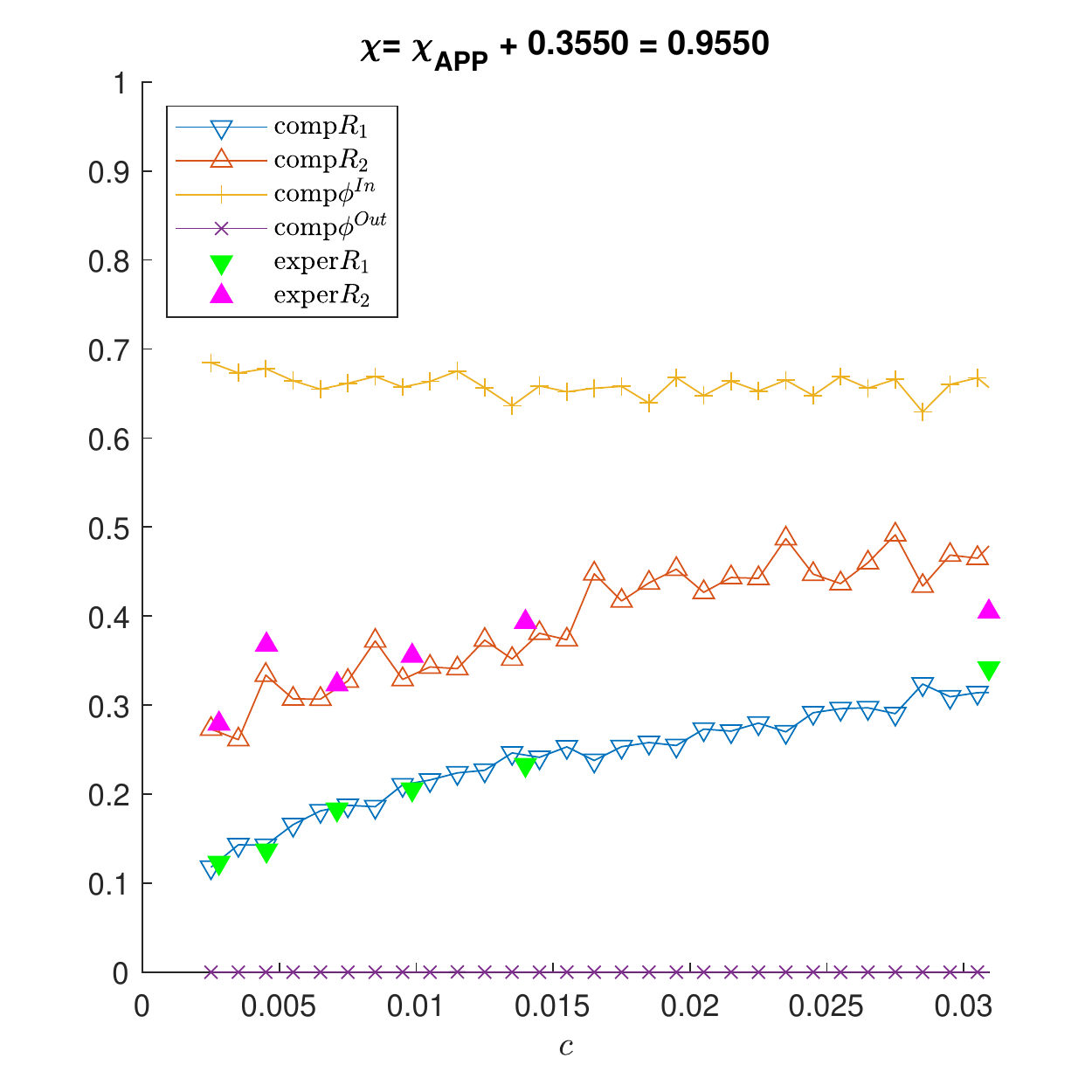}
\caption{Computed vs. experimental values of the torus radii, $R_1$ and $R_2$, as well as the volume fractions $\phi^{in}$ and $\phi^{out}$, for $\chi=\ca+0.32 = 0.92$ (left) and $\chi=\ca+0.355= 0.955$ (right). Other parameters are taken to be constant: $ \gamma = 2.8956,\,N_1 = 100, \,N_2 = 1, \, \eta = 1000$.}
\label{fig:PSO}
\end{figure}

To compare to the experimental data, and since we do not have experimental values of $\phi^{in}$ and $\phi^{out}$, we calculate the values of $c$ using the formula in (\ref{MainConstraints1}), where we use experimental values for $R_1$ and $R_2$ and average values of the computed $\phi^{in}$ and $\phi^{out}$. Using the particle swarm optimization method, we search within the intervals $[0.1, 0.5],\: [0.2, 0.7],\: [0.5, 0.7],$ and $[0, 0.1]$ for $R_1, R_2, \phi^{in}$, and $ \phi^{out}$, respectively. In addition to the constraint (\ref{MainConstraints1}), we enforce the last two geometric constraints in (\ref{MainConstraints2}). We use 1000 particles and 50 generations for each simulation. Figure \ref{fig:PSO} visualizes the computed versus experimental solutions for specific values of $\chi$ greater than the critical value.

We conclude by listing several observations from the simulations.
\begin{list}{$\circ$}{}  
	\item For any value of $\chi$, the computation of $R_1$ seems to be more accurate than that of $R_2$, as in the reduced model. 
	\item For any value of $\chi$, the value of $\phi^{out}$ is negligible and the value of $\phi^{in}$ is almost constant.
	\item The range of values of $\chi$ that seem to work best is $\ca+[0.31, 0.37]$, which means that $\chi$ ranges between $0.91$ and $0.97$. 
\end{list}	

\section{Conclusions}\label{sec:conclusion}
We have studied the aggregation phenomena in lyotropic chromonic liquid crystals, justifying toroidal shapes observed and simulating torus sizes for different types of experiments. Follow-up work would address some of the simplifying assumptions made in the present paper. First of all, including chirality in the Oseen-Frank energy will allow modeling of cholesteric shapes, providing a better tool 
to study DNA aggregates. The work would also benefit from considering shapes more general than the torus, by generalizing the free boundary problem allowing for axisymmetric domains (in the non-chiral case) and helicoid shapes in the cholesteric one. Finally, studying the time evolution of the clusters to equilibrium would shed some light on the metastable features of the phenomena. 

\medskip
\section{Appendix: The bending energy of a torus}\label{appendix}
Here we write the bending energy specific for the case when $\Omega$ is a torus. Basically, we show how to go from Equation~\eqref{bending_energy_2} to the expression corresponding to the bending energy in \eqref{eq: Torus Energy}, i.e.,
$$E_{b}=\kappa _3 R_0\int _{{\rm Torus}}\phi\left(|(\nabla \times \mathbf{n})\times \mathbf{n}|^2\right)d\bar{\mathbf{x}} = 4\pi^2 \kappa _3 \phi \left(R_2-\sqrt{R_2^2-R_1^2}\right). $$
Recall that for the torus, we have that $\mathbf{n}=\left(-\frac{y}{r}, \frac{x}{r},0\right)$ with $r = \sqrt{x^2+y^2}$. Note that $\frac{\partial r}{\partial x} = \frac{x}{r}$ and $\frac{\partial r}{\partial y} = \frac{y}{r}$. Then, we have that 
\begin{align}
\nabla \times \mathbf{n} &=
\left| \begin{array}{ccc}
\mathbf{i} & \mathbf{j} &  \mathbf{k}  \\ 
\partial_x & \partial_y & \partial_z  \nonumber \\ 
\frac{y}{r} & \frac{x}{y} & 0
\end{array} \right| = \left( \frac{\partial}{\partial x}\frac{x}{r} + \frac{\partial}{\partial y}\frac{y}{r} \right) \mathbf{k}   
= \left( \frac{r-x\frac{x}{r}}{r^2} + \frac{r-y\frac{y}{r}}{r^2} \right) \mathbf{k} \nonumber\\
& = \left( \frac{r^2-x^2}{r^3} + \frac{r^2-y^2}{r^3} \right) \mathbf{k}  = \left( \frac{x^2+y^2}{r^3} \right) \mathbf{k} = \frac{r^2}{r^3} \mathbf{k} = \frac{1}{r} \mathbf{k},\nonumber  
\end{align}
which give us the expression
$$
\mathbf{n} \times (\nabla \times \mathbf{n}) =
\left| \begin{array}{ccc}
\mathbf{i} & \mathbf{j} &  \mathbf{k}  \\ 
-\frac{y}{r} & \frac{x}{r} & 0  \\ 
0 & 0 & \frac{1}{r}
\end{array} \right| = \left( \frac{x}{r^2} \right) \mathbf{i} - \left( \frac{y}{r^2} \right) \mathbf{j},$$
and finally
$$ |\mathbf{n} \times (\nabla \times \mathbf{n})|^2 = \frac{x^2+y^2}{r^4} = \frac{1}{r^2}.$$

Thus, 
\begin{equation}\label{eq: Eb_torus}
E_{b}=\kappa_3 R_0\int _{{\rm Torus}}\phi \frac{1}{r^2} dV = \kappa_3 R_0\phi \int _{{\rm Torus}} \frac{1}{r^2} dV.
\end{equation}

\begin{figure}[htb!]
    \centering
 \includegraphics[scale=0.4]{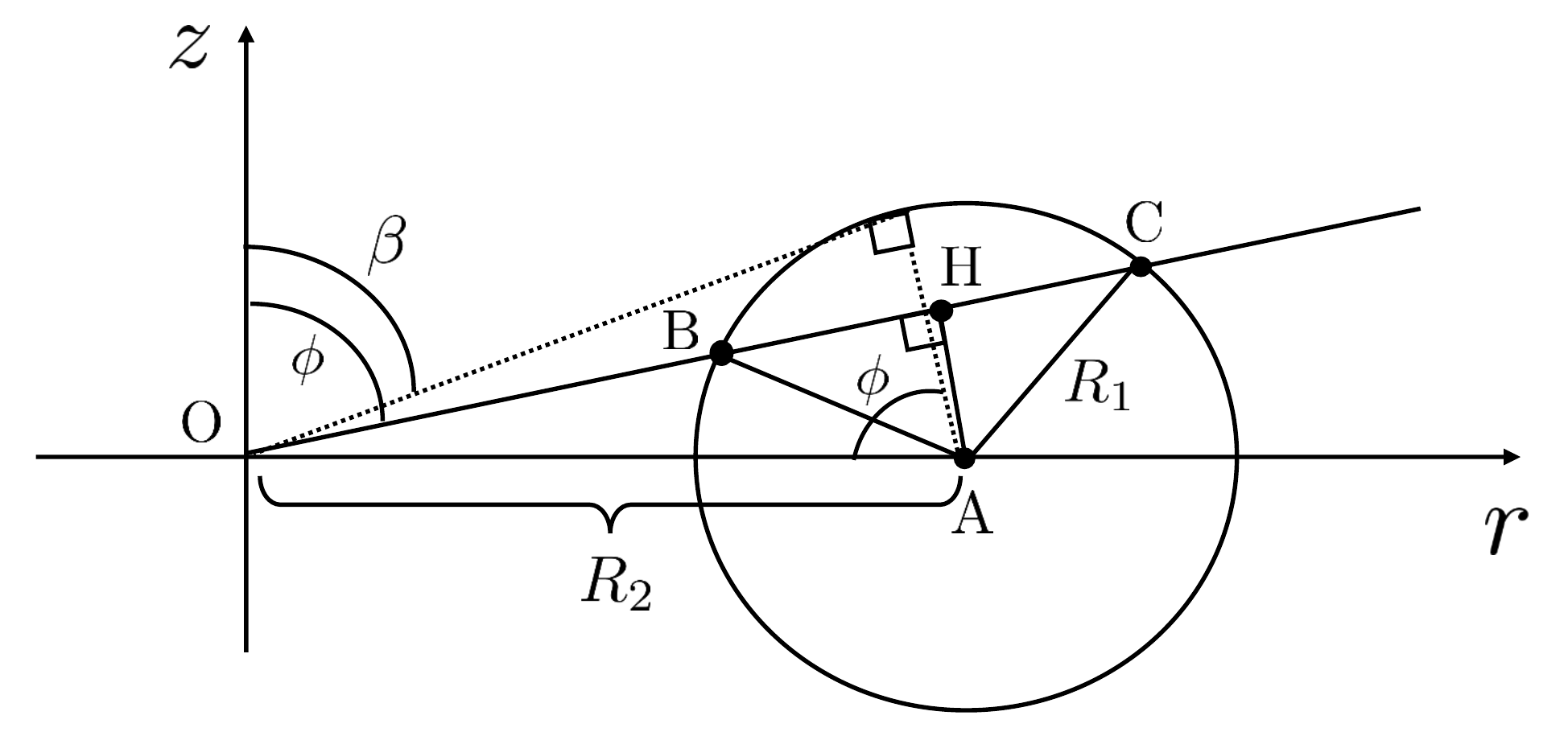}
  \caption{Diagram with spherical coordinates}
    \label{fig:spherical}
\end{figure}

To find this last integral we shall use spherical coordinates as done in \cite{torusvolume}. Following the diagram in Figure~\ref{fig:spherical}, we define the segment lengths $\ell_1$=OB and $\ell_2$=OC, the angle $\phi$ formed by the $z$-axis and the segment OC. In the triangle OHA, we have that the length of OA is equal to $R_2$ and the angle $\widehat{{\rm OAH}}=\phi$. We can then write that ${\rm OH}=R_2\sin{\phi}$ and ${\rm AH}=R_2\cos \phi$. In the triangle ABC,  we have that BH$^2=$AB$^2-$AH$^2$=$R_1^2-R_2^2\cos^2\phi$. Then, we can write
\begin{align}\ell_1={\rm OB}={\rm OH-BH}=R_2\sin{\phi}-\sqrt{R_1^2-R_2^2\cos^2 \phi} \nonumber\\
\ell_2={\rm OC}={\rm OH + BH}=R_2\sin{\phi} + \sqrt{R_1^2-R_2^2\cos^2 \phi} \nonumber
\end{align}
and reformulate the integral in \eqref{eq: Eb_torus}.

\begin{align}
    \int _{{\rm Torus}} \frac{1}{r^2} dV & = \int_0^{2\pi} \int_\beta^{\pi-\beta}\int_{\ell_1}^{\ell_2} \frac{1}{\ell^2 \sin^2 \phi} \ell^2 \sin \phi \, d\ell \,  d\phi \, d\theta  \nonumber \\
    & = \int_0^{2\pi} \int_\beta^{\pi-\beta} \frac{\ell_2-\ell_1}{\sin \phi} \,  d\phi \, d\theta  
    = 2 \pi \int_\beta^{\pi-\beta} \frac{2\sqrt{R_1^2-R_2^2\cos^2\phi}}{\sin \phi} \,  d\phi \nonumber \\
    & = 4 \pi R_1 \int_{-1}^{1} \frac{\sqrt{1-u^2}}{\sin \phi} \left(- \frac{R_1}{R_2 \sin \phi}\right)  \,  d\phi \quad (\mbox{with } u = \frac{R_2}{R_1}\cos \phi)  \nonumber \\
    & = 4 \pi \frac{R_1^2}{R_2} \int_{-1}^{1} \frac{\sqrt{1-u^2}}{1 - \frac{R_1^2}{R_2^2}u^2} \,  du 
    = 4 \pi R_2 \int_{-1}^{1} \frac{\sqrt{1-u^2}}{\lambda^2 - u^2} \,  du \quad (\mbox{with } \lambda = \frac{R_2}{R_1})\nonumber \\
    & = 4 \pi R_2 \int_{-\pi/2}^{\pi/2} \frac{\cos \alpha}{\lambda^2 - \sin^2 \alpha} \cos \alpha \,  d\alpha \quad  \left(\mbox{with } u = \sin \alpha\right) \nonumber \\
    & = 4 \pi R_2 \int_{-\pi/2}^{\pi/2} \frac{1-\sin^2 \alpha}{\lambda^2 - \sin^2 \alpha}\,  d\alpha 
    = 4 \pi R_2 \int_{-\pi/2}^{\pi/2} \frac{1-\sin^2 \alpha + \lambda^2 - \lambda^2}{\lambda^2 - \sin^2 \alpha}\,  d\alpha \nonumber \\
    & = 4 \pi R_2 \left\{\int_{-\pi/2}^{\pi/2} 1 \, d\alpha +  (1-\lambda^2)\int_{-\pi/2}^{\pi/2} \frac{1}{\lambda^2 - \sin^2 \alpha}\,  d\alpha \right\} \nonumber \\
    & = 4 \pi R_2 \left\{\pi +  \frac{1-\lambda^2}{\lambda^2}\int_{-\pi/2}^{\pi/2} \frac{1}{1 - \frac{1}{\lambda^2}\sin^2 \alpha}\,  d\alpha \right\} \nonumber \\
    & = 4 \pi R_2 \left\{\pi +  \frac{1-\lambda^2}{\lambda^2}\int_{-\pi/2}^{\pi/2} \frac{1}{\cos^2\alpha + \sin^2 \alpha - \frac{1}{\lambda^2}\sin^2 \alpha}\,  d\alpha \right\} \nonumber \\
    & = 4 \pi R_2 \left\{\pi +  \frac{1-\lambda^2}{\lambda^2}\int_{-\pi/2}^{\pi/2} \frac{1}{\cos^2\alpha +  \left(\frac{\lambda^2-1}{\lambda^2}\right)\sin^2 \alpha}\,  d\alpha \right\} \nonumber \\
    & = 4 \pi R_2 \left\{\pi +  \frac{1-\lambda^2}{\lambda^2}\int_{-\pi/2}^{\pi/2} \frac{1}{1 +  \left(\frac{\lambda^2-1}{\lambda^2}\right)\tan^2 \alpha} \frac{1}{\cos^2 \alpha}\,  d\alpha \right\} \nonumber \\
    & = 4 \pi R_2 \left\{\pi +  \frac{1-\lambda^2}{\lambda^2}\frac{\lambda}{\sqrt{\lambda^2-1}} \int_{-\infty}^{\infty} \frac{1}{1 + v^2}  \,  dv \right\}(\mbox{with } v = \frac{\sqrt{\lambda^2-1}}{\lambda}\tan \alpha) \nonumber \\
    & = 4 \pi R_2 \left\{\pi - \pi \frac{\sqrt{\lambda^2-1}}{\lambda} \right\} 
    = 4 \pi R_2 \left\{\pi - \pi \frac{\sqrt{\frac{R_2^2}{R_1^2}-1}}{\frac{R_2}{R_1}} \right\} 
    = 4 \pi^2 \left(R_2 - \sqrt{R_2^2-R_1^2}\right). \nonumber
\end{align}

\section*{Acknowledgements}
M. Carme Calderer acknowledges the support from the National Science Foundation, grant number DMS-DMREF 1729589 and DMS-1816740. The authors are grateful to the support of the Michigan Center for Applied and Interdisciplinary Mathematics at the University of Michigan, for hosting the Women in Mathematics of Materials (WIMM) Workshop that allowed this research to get underway and the partial support provided by the Association for Women in Mathematics (AWM) through the AWM-ADVANCE grant NSF-HRD 1500481. The authors thank Runa Koizumi for helpful discussions regarding the experimental data, and the reviewers for their thoughtful comments and effort towards improving the manuscript.

\end{document}